\documentclass[12pt]{article}

\usepackage{setspace}
\usepackage[utf8]{inputenc}
\usepackage{amssymb}
\usepackage{amsmath}
\usepackage{amsthm}
\usepackage{geometry}
\usepackage{graphicx}
\usepackage[export]{adjustbox}
\usepackage{wrapfig}
\usepackage{float}
\usepackage{mathrsfs}
\usepackage[colorlinks=true, allcolors=blue]{hyperref}
\usepackage{relsize}
\usepackage[shortlabels]{enumitem}
\usepackage{setspace}
\usepackage{multirow}
\usepackage{subfig}
\usepackage{bbm}
\usepackage[title]{appendix}
\usepackage{tikz}
\usepackage{calc}
\usepackage{bbm}
\usepackage{natbib}

\usetikzlibrary{graphs, graphs.standard}
\usetikzlibrary{decorations.text,decorations.markings}

\usepackage{color} 
\usepackage{listings} 
\usepackage{setspace}
\usepackage{pgfplots}
 
\definecolor{Code}{rgb}{0,0,0} 
\definecolor{Decorators}{rgb}{0.5,0.5,0.5} 
\definecolor{Numbers}{rgb}{0.5,0,0} 
\definecolor{MatchingBrackets}{rgb}{0.25,0.5,0.5} 
\definecolor{Keywords}{rgb}{0,0,1} 
\definecolor{self}{rgb}{0,0,0} 
\definecolor{Strings}{rgb}{0,0.63,0} 
\definecolor{Comments}{rgb}{0,0.63,1} 
\definecolor{Backquotes}{rgb}{0,0,0} 
\definecolor{Classname}{rgb}{0,0,0} 
\definecolor{FunctionName}{rgb}{0,0,0} 
\definecolor{Operators}{rgb}{0,0,0} 
\definecolor{Background}{rgb}{0.98,0.98,0.98} 

\newtheorem{theorem}{Theorem}
\newtheorem{proposition}{Proposition}
\newtheorem{assumption}{Assumption}

\newcommand{\hbAppendixPrefix}{A.}

\newtheorem{lemma}{Lemma}

\theoremstyle{definition}
\newtheorem{definition}{Definition}

\newtheorem{examp}{Example}
\newenvironment{example}{\begin{examp}}{\mbox{} \nolinebreak \hfill \mbox{$\Diamond$} \end{examp}}

\DeclareMathOperator*{\argmax}{argmax}
\DeclareMathOperator*{\argmin}{argmin}

\usepackage{color,soul}

\begin{document}
\date{April 2025}
\author{Jose M. Betancourt\footnote{\noindent Email: jose.betancourtvalencia@yale.edu. \\
I am grateful to Ben Golub, Tomás Rodríguez-Barraquer, Simon Boutin, Angelo Mele, Elliot Lipnowski, Doron Ravid, Aniko Öry, Kevin Williams, Kevin Yin, Annie Chen, Laura Tenjo, and Eric Solomon for very informative discussions. I thank seminar participants at the Network Science and Economics Conference for helpful comments and suggestions.
}\\
\small{Yale University}
} 
\title{{\bf Tractability and Phase Transitions in Endogenous Network Formation}
}

\maketitle

\begin{abstract}
    The dynamics of network formation are generally very complex, making the study of distributions over the space of networks often intractable. Under a condition called \textit{conservativeness}, I show that the stationary distribution of a network formation process can be found in closed form, and is given by a Gibbs measure. For conservative processes, the stationary distribution of a certain class of models can be characterized for an arbitrarily large number of players. In this limit, the statistical properties of the model can exhibit \textit{phase transitions}: discontinuous changes as a response to continuous changes in model parameters.
    
\end{abstract}

\clearpage

\section{Introduction}
The structure of which economic agents interact with each other -- be it individuals, firms, countries, etc -- is central in determining the outcomes of these interactions. Describing in a tractable way how these structures are formed can provide insights into the dynamics of these economic processes. A fundamental problem in studying network formation models is that the number of networks increases exponentially in the number of nodes, making it difficult to characterize the properties of the resulting networks when there are many players. In this paper, I propose a theoretical model of endogenous network formation which allows for a tractable characterization even for arbitrarily large networks.

The process of forming a network often involves a combination of an element of randomness and strategic decisions. For example, one might randomly meet a friend of a friend at a social gathering, but the decision of whether or not to develop the relationship after this initial meeting is a conscious choice. Such interactions and choices happen constantly, making networks highly dynamic and fluid objects.

A model that incorporates random meetings and strategic decisions will give rise to a probability distribution over the space of networks that evolves through time. The goal of this paper is to analyze the properties of such a distribution. Specifically, I study whether the stationary distribution of networks can be written as a function of the utilities that drive the network formation decisions. This corresponds to characterizing the long-run properties of the system in terms of the model primitives.

Even by simplifying the problem to the study of stationary distributions, finding a closed-form expression for the stationary probabilities is very computationally expensive. Intuitively, this is because there is a transition rate for every network and every possible set of two agents. This means that there are around $N^2 2^{N^2}$ transition rates, and the stationary distribution is a function of the full set of transition rates.

The first main result of the paper is that the stationary distribution of the network formation process can be written in closed form if individuals' utility functions have a special structure, which I call \textit{conservativeness}. This condition corresponds to two equivalent properties of the process. The first is that the network formation process is a reversible Markov chain, which allows for the closed-form expression of the stationary distribution. The second property is that, in the absence of randomness, the network formation process is a potential game. Both properties represent some reduction in the dimensionality of the system by exploiting a symmetry condition.

If utility functions satisfy conservativeness, the stationary distribution is a Gibbs measure. Specifically, the stationary probability associated to a network $g$ is proportional to $\exp(\Phi(g))$, where $\Phi(g)$ is the potential associated to $g$ from the game with no randomness \citep{Monderer1996PotentialGames, Chakrabarti2007NetworkPotentials}. This yields an explicit connection between the stochastic network formation process and the corresponding deterministic network formation game.

The second main result of the paper concerns the structure of conservative utility functions. Two sets of utility functions are said to be \textit{choice-equivalent} if they yield the same network formation process. The result is that any set of utility functions that is conservative is choice-equivalent to one where individuals derive value from being part of group structures. Specifically, individuals assign value to every possible network $g_0$, and if $g_0$ is a subnetwork of $g$, then all individuals that are nodes of $g_0$ obtain the same utility. This means that conservativeness is tightly linked to the value of groups, which arises from it being a symmetry condition.

The equivalence between conservativeness and assigning value to group structure yields a natural scheme to scale up the number of players, allowing for the analysis of arbitrarily large networks. The scheme consists of individuals deriving utility from a fixed set of networks called \textit{motifs}. Every time a motif is realized in a network, individuals corresponding to the nodes get the value assigned to that motif. Under this scheme, we can study the dynamics of the model for an arbitrary number of players by specifying a finite set of motifs. The simplicity of this scheme allows for a simple characterization of the asymptotic properties of the process as the number of players becomes large.

The last main result of the paper is that the model with motif utilities becomes equivalent (in an appropriate graph limit sense) to an Erd\"os-R\'enyi model, where the parameter of the model is determined by an optimization process. The result follows from the literature on large deviations theory applied to large dense graphs \citep{Chatterjee2013}. This optimization problem corresponds to a trade-off between agents' utilities and a quantity called \textit{entropy}, which takes into account the exponentially increasing size of network space. Introducing multiple types of agents generalizes this result, such that the process instead converges to a stochastic block model. In this case, the linking probabilities are determined by an analogous optimization process. 

The nature of the optimization process allows for the emergence of \textit{phase transitions} in the asymptotic properties of the process. Since the parameter in the equivalent Erd\"os-R\'enyi model is the minimum of some objective function, even a continuous change in the model parameters can lead to a discontinuous change in this minimum. These discrete jumps are referred to as phase transitions, since the network discretely changes from a high-density phase to a low-density phase and vice versa.

To illustrate how these phase transitions might arise naturally in an economic setting, I study a simple model of trade, where firms stochastically meet and decide whether to form trade relationships. Firms must incur in some distance-dependent cost to state the intent to trade with another firm, and they derive value when there is a mutual intent to trade. If the benefits from trade are taxed, increasing the tax rate can induce a phase transition, breaking the trading network apart. More complex trading structures lead to diverse manifestations of phase transitions, illustrating the richness of this framework.

Together with the results on the closed-form expression of the stationary distribution of the process, the results on phase transitions shed light on the very complex process of network formation. The low-dimensional behavior of the asymptotics allow for potential applications of this framework to more complex macroeconomic models, which take into account the details of network formation. These results showcase novel phenomena that are relevant for the dynamics of and interventions on complex network structures.

\subsection{Related Literature}
This paper contributes to the broad literature on the analysis of endogenous network formation models. A common feature of these models is to include stochasticity in the meeting process and agents' utilities to obtain tractable results. See \citet{Jackson1996, Bala2000AFormation, Jackson2001TheNetworks, Jackson2012SocialExchange} for examples of deterministic network formation and \citet{Jackson2002TheNetworks, Currarini2009AnMinoritiesandSegregation, golub2010strategic, Mele2017AFormation, mele2022structural} for stochastic network formation. See \citet{jackson2005survey, jackson2008social} for more complete surveys of network formation models and their properties.

Recent work on network formation includes additional choices in addition to link formation/deletion and analyzes the resulting network, such as \citet{hsieh2022structural, sadler2024gamesendogenousnetworks}. \citet{badev2021nash} considers simultaneous changes to the state of multiple network links. This paper differs from these in that more general utility functions for network formation are considered, but only in settings with single link evaluations and with no additional actions. For the case of forward-looking agents, \citet{dutta2005farsighted} establishes equilibrium existence results in a more general setting, but lacks a tractable characterization of the resulting probabilities over the space of networks.

A special class of models that is of interest in econometric estimation are Exponential Random Graph Models (ERGMs)\footnote{See \citet{robins2007introduction} for an overview of ERGMs}. In these models, the probability associated to network $g$ is proportional to $\exp(\Phi(g))$, where $\Phi(g)$ is a linear combination of \textit{sufficient statistics} of the network. The advantage of analyzing these models is twofold. First, it has been shown that they arise naturally from strategic network formation models under some regularity assumptions, which would otherwise be intractable \citep{Butts2009UsingModels, chandrasekhar2014tractable, Mele2017AFormation, christakis2020empirical}. The second advantage of ERGMs is that they provide a natural tool to analyze large graphs, provided that the sufficient statistics scale properly as the number of agents grows \citep{aristoff2017phase, Mele2017AFormation, mele2022structural}. This paper contributes to the literature on ERGMs by characterizing a model of network formation that provides a microfoundation for distributions with arbitrary functions $\Phi(g)$, and hence any ERGM. The model can then be evaluated to see if the primitives satisfy desirable properties, which would yield criteria to evaluate the use of a given ERGM.

Finally, this paper contributes to the literature on graph limits by providing a characterization of the limiting properties of large networks in terms of model primitives. In the case of ERGMs, the distributions can be characterized in the limit of a large number of players, since the model becomes intractable when this number is large but finite\footnote{Previous work on tackling the intractability of these models includes \citet{boucher2017my}, which uses Markov Random Fields to reduce the complexity in estimation, and \citet{graham2017econometric}, which proposes consistent estimators of structural parameters that take into account degree heterogeneity. \citet{Mele2017AFormation} and \citet{mele2022structural} use an exchange Markov chain Monte Carlo method to estimate parameters through maximum likelihood in ERGMs.}. Results from the theory of graph limits \citep{chatterjee2011large, Chatterjee2013, Mele2017AFormation} allow for a characterization of the limiting behavior of the process. This characterization makes it evident that the model can exhibit phase transitions\footnote{Phase transitions are a central topic of analysis in statistical physics. See, for example, \citet{Holyst2000PhaseFormation, dorogovtsev2008critical, squartini2015breaking, cimini2019statistical} for applications of statistical physics to stochastic network analysis.}, a phenomenon that has been observed in the context of social network formation \citep{golub2010strategic} and supply chain formation \citep{elliott2022supply}. 
\section{Myopic Network Formation} \label{sec:model}

\subsection{Stochastic meeting of agents}
Consider a set ${\cal N} = \{1,\ldots,N\}$ of $N$ individuals who can interact with each other on a directed unweighted graph\footnote{The results can be generalized to undirected networks if we allow for transfers when agents meet each other.}. I use the notation $ij$ to refer to the tuple $(i,j)$. Since I do not consider self-interactions, the set of possible dyads associated to these individuals is given by ${\cal D}=\{ij \in {\cal N}^2 \, | \, i \ne j\}$. A network $g$ on ${\cal N}$ is defined as a subset of ${\cal D}$, corresponding to the dyads for which there is a link. Let ${\cal G} = {\cal P}({\cal D})$, where ${\cal P}$ denotes the power set, be the set of all possible networks on ${\cal N}$. 

In order to describe the process of link updating, it will be useful to define a set of ``switching'' functions. Consider the functions $(\sigma_{ij})_{ij \in {\cal D}}$ defined by
\begin{align}
    \sigma_{ij}(g) =
    \begin{cases}
        g \cup \{ij\} & \textrm{if } ij \not \in g \\
        g \backslash \{ij\} & \textrm{if } ij \in g.
    \end{cases}
\end{align}
Intuitively, the operation $\sigma_{ij}$ creates the link $ij$ if it is not present in a network and severs it if it is.

The network formation process occurs in continuous time. Agents meet stochastically and, conditional on meeting, decide whether they want to change the state of the network. If the network state is $g$, agent $i$ meets agent $j$ at an exogenous Poisson rate $\lambda_{ij}(g)$, that can potentially depend on $g$. Following \cite{Mele2017AFormation}, I impose the following restriction on the meeting rates:
\begin{assumption} \label{A:meeting_rates}
    The meeting rates satisfy $\lambda_{ij}(g) = \lambda_{ij}(\sigma_{ij}(g))$ for all dyads $ij$ and all networks $g$.
\end{assumption}
Assumption \ref{A:meeting_rates} means that the meeting rate for the dyad $ij$ is independent of whether it is present in the network, but can depend on the rest of the rest of the network.

Conditional on meeting $j$, agent $i$ makes a decision to keep or change the state of their relationship through a discrete choice rule, parametrized by a set of utility functions $V: {\cal N} \times {\cal G} \to \mathbb{R}$. Agent $i$ has an baseline utility $V_i(g)$ associated to network $g$, and there are idiosyncratic shocks to this utility every time a choice is made. The utility agent $i$ obtains from changing the state of their link to agent $j$ given a network $g$ is given by $u_{ij}^1(g)$, while the utility of not changing the state of the link is $u_{ij}^0(g)$. In terms of the baseline utilities, these are given by
\begin{align} \label{eq:disc_choice}
    u_{ij}^0(g) &= V_i(g) + \varepsilon_{ij}^0, \nonumber \\
    u_{ij}^1(g) &= V_i(\sigma_{ij}(g)) + \varepsilon_{ij}^1,
\end{align}
where the shocks $\varepsilon_{ij}^k$ are i.i.d. and follow some distribution $F_0$. Let $F_1$ be the distribution of $\varepsilon^1_{ij} - \varepsilon^0_{ij}$, such that the probability of the network changing from $g$ to $\sigma_{ij}(g)$, denoted by $p_{ij}(g)$, is given by
\begin{align}
    p_{ij}(g) = F_1(V_i(\sigma_{ij}(g)) - V_i(g)).
\end{align}
With this characterization, we can now analyze how the dynamics of the model depend on the structure of the utility functions.

\subsection{Dynamics and convergence}
I assume at time $t=0$ there is some initial probability distribution $\pi_0 \in \Delta({\cal G})$ over the space of networks. Let $G_t$ be the realization of a network at time $t$ and let $\pi_t(g)$ be the probability measure associated to $g$ at time $t$. The measure $\pi_t$ evolves according to the Kolmogorov forward equation:
\begin{align}
    \dot{\pi}_t(g) = \sum_{ij \in {\cal D}} \lambda_{ij}(g) \left[p_{ij}(\sigma_{ij}(g)) \pi_t(\sigma_{ij}(g)) - p_{ij}(g) \pi_t(g) \right].
\end{align}
Note that, as long as the distribution of utility shocks $F_0$ has full support, the switching probabilities $p_{ij}(g)$ are non-degenerate. Therefore, in this case, the network formation process corresponds to an aperiodic irreducible Markov process on ${\cal G}$. It follows from standard results in stochastic processes \citep{Rosenthal2006ATheory} that there is a unique distribution $\pi$ such that $\pi_t(g) \to \pi(g)$ for all $g \in {\cal G}$, regardless of the initial conditions. Assumption \ref{A:T1EV}, which permits tractability of the analysis to follow, ensures that this is the case.
\begin{assumption} \label{A:T1EV}
    The shocks $\varepsilon_{ij}^k$ follow a type 1 extreme value (T1EV) distribution.
\end{assumption}

Assumption \ref{A:T1EV} implies that the switching probabilities $p_{ij}(g)$ are a logistic function of the marginal utility of switching $V_i(\sigma_{ij}(g)) - V_i(g)$. Equivalently, they satisfy
\begin{align} \label{eq:log_odds_ratio}
    \log \left( \frac{p_{ij}(g)}{p_{ij}(\sigma_{ij}(g))} \right) = V_i(\sigma_{ij}(g)) - V_i(g).
\end{align}
This yields an intuitive connection between the properties of the Markov chain (through the transition probabilities) and the utility functions of the agents.

Given that the dynamics of the network formation process are related in a simple way to the utility functions of the agents, we would expect that reducing the complexity of the incentive structure will also reduce the complexity of the network formation dynamics. One way to do this is by imposing a ``symmetry'' condition on the actions of agents, as specified in Definition \ref{def:conservative}.
\begin{definition} \label{def:conservative}
    A set of utility functions $V: {\cal N} \times {\cal G} \to \mathbb{R}$ is said to be \textit{conservative} if the following condition on marginal utilities (MUs) holds:
    \begin{align} \label{eq:conservative_utility}
        &\underbrace{V_i(\sigma_{ij}(g)) - V_i(g)}_{\textrm{MU of changing link $ij$ in network $g$}} + \underbrace{V_{i'}(\sigma_{i'j'} \circ \sigma_{ij}(g)) - V_{i'}(\sigma_{ij}(g))}_{\textrm{MU of changing link $i'j'$ in network $\sigma_{ij}(g)$}} \nonumber \\
        &= \underbrace{V_{i'}(\sigma_{i'j'}(g)) - V_{i'}(g)}_{\textrm{MU of changing link $i'j'$ in network $g$}} + \underbrace{V_{i}(\sigma_{ij} \circ \sigma_{i'j'}(g)) - V_{i}(\sigma_{i'j'}(g))}_{\textrm{MU of changing link $ij$ in network $\sigma_{i'j'}(g)$}}
    \end{align}
    for all $g \in {\cal G}$ and all $ij, i'j' \in {\cal D}$. That is, the sum of the marginal utilities of changing the state of links $ij$ and $i'j'$ is independent of the order in which the changes are made.
\end{definition}
Utility functions satisfying the property in Definition \ref{def:conservative} are called conservative in analogy to conservative vector fields\footnote{A $C^1$ vector field ${\bf F}:U \to \mathbb{R}^n$, with $U \subset \mathbb{R}^n$ open and simply connected, is said to be conservative if there exists a function $\phi$ such that ${\bf F}({\bf x}) = \nabla \phi({\bf x})$. According to the Poincar\'e Lemma, if $\forall i,j \in \{1,\ldots,n\}$, $\frac{\partial F_i({\bf x})}{\partial x_j} = \frac{\partial F_j({\bf x})}{\partial x_i}$, then ${\bf F}$ is conservative \citep{warner1983foundations}.\label{foot:conservative}}. The condition in Equation \eqref{eq:conservative_utility} can be thought of as the condition for the Poincar\'e Lemma in Footnote \ref{foot:conservative}, which allows for the characterization of a vector field as the gradient of a scalar field, effectively reducing the dimensionality of the object. Proposition \ref{prop:conv} formalizes this intuition, showing that this condition is necessary and sufficient to reduce the complexity of the incentive structure and of the network formation process, in a suitable manner.

\begin{proposition} \label{prop:conv}
The following are equivalent:
\begin{enumerate}
    \item The utility functions are conservative.
    \item There exists a function $\Phi:{\cal G} \to \mathbb{R}$, unique up to an additive constant, such that
    \begin{align}
        \Phi(\sigma_{ij}(g)) - \Phi(g) = V_i(\sigma_{ij}(g)) - V_i(g).
    \end{align}
    \item The network formation process is reversible.
\end{enumerate}
Furthermore, if the utility functions are conservative, the stationary distribution of the process is given by the Gibbs measure
\begin{align} \label{eq:eq_dist}
    \pi(g) = \frac{\exp(\Phi(g))}{\sum_{g' \in {\cal G}} \exp(\Phi(g'))},
\end{align}
where $\Phi$ is the function in statement 2.
\end{proposition}

Proposition \ref{prop:conv} states that conservativeness of the utility functions yields three properties of the network formation process that reduce the dimensionality of the problem. First, we see that the information about the incentive structure of switching the states of links can be encoded in a common function $\Phi$. This is called a \textit{potential function} \citep{Monderer1996PotentialGames}, and it has properties that make it useful to simplify the analysis of complex games, such as characterizing the Nash equilibria as local maxima of the potential\footnote{The existence of a potential arising from the conservativeness condition is analogous to Theorem 2.8 in \citet{Monderer1996PotentialGames}.}. Second, given the simple connection between the switching probabilities and the utility functions, conservativeness is also related to the reversibility of the Markov process\footnote{Reversibility of a Markov jump process on a finite space ${\cal X}$ is usually stated in terms of the detailed balance condition: $W(i,j) \pi(i) = W(j,i) \pi(j)$ for all $i,j \in {\cal X}$, where $W(\cdot)$ are the transition rates and $\pi(\cdot)$ are the stationary probabilities. An equivalent characterization solely in terms of rates is Kolmogorov's criterion, which is analogous to conservativeness \citep{Norris1997Continuous-timeII}.}.  Reversibility implies the third property, which is that there exists a function that encodes the transition rates of the stochastic process\footnote{In the theory of Markov processes, this function is also called a potential, or energy, function.}. Interestingly, it is precisely the potential function of the game that determines the stationary distribution of the network formation process.

This characterization of the stationary distribution is related to the family of exponential random graph models (ERGMs). These assign a probability to a network of the form
\begin{align} \label{eq:ERGM}
    \pi_{\textrm{ERGM}}(g) = \frac{\exp(\boldsymbol{\beta} \cdot {\bf S}(g))}{\sum_{g' \in {\cal G}} \exp(\boldsymbol{\beta} \cdot {\bf S}(g'))},
\end{align}
where ${\bf S}(g)$ is a vector of sufficient statistics of the network and $\boldsymbol{\beta}$ is a vector of model parameters. Therefore, we see that if the potential of our network formation game is of the form $\Phi(g) = \boldsymbol{\beta} \cdot {\bf S}(g)$, the stationary distribution will be the distribution of an ERGM with sufficient statistics ${\bf S}$. This means that this myopic network formation model serves as a microfoundation for ERGMs, conditional on finding suitable utility functions\footnote{\citet{Mele2017AFormation} provides a microfoundation for a class of ERGMs with this model. \citet{chandrasekhar2014tractable} generalizes this analysis to a wider class of utility functions.}. One property of ERGMs that is shared by the myopic network formation game is that the denominator in the Gibbs measure involves a sum over all possible networks, making its calculation intractable. This intractability causes problems for estimation of ERGMs (see \citet{chandrasekhar2014tractable} for a detailed description of the problems with estimating ERGMs and potential solutions). However, as I will discuss in Section \ref{sec:large_networks}, for a certain class of models where the number of agents can be systematically increased, characterizing the large $N$ behavior of this denominator allows for even further dimensionality reduction of the model.

Note that Proposition \ref{prop:conv} states whether a given set of utility functions can be represented in terms of a potential, and we can construct the potential if this is the case (see the proof in the Appendix). However, it does not state how to determine which utility functions can generate a given potential. A trivial way to do this is by having all agents have the same utility function $V_i(g) = \Phi(g)$. While this yields a Markov process that samples from the distribution in Equation \eqref{eq:eq_dist} (in fact, it corresponds to the Glauber algorithm \citep{Glauber2004TimeDependentModel}), it removes all the heterogeneity in agents' preferences. I now analyze how the structure of the stationary distribution can be preserved while allowing for heterogeneity in preferences.

\subsection{Equivalence of utilities}
In the previous section I showed conditions on the utility functions under which the stationary distribution of the network formation process can be expressed in terms of the potential of a game. Clearly this representation is not unique, as can be easily seen by shifting the utility functions by some agent-dependent constant: $V_i(g) \to V_i(g) + c_i$. If the original game had an associated potential $\Phi$, this will also be a potential of the modified game, and that the dynamics of the model will remain unchanged. Particularly, if the dynamics of the model remain unchanged under a transformation, the associated potential of the game will not change.

\begin{definition}
    Two sets of utility functions $V$ and $V'$ are said to be \textit{choice-equivalent} if the switching probabilities $p_{ij}(g)$ are the same under $V$ and $V'$ for all dyads $ij$ and networks $g$.
\end{definition}

Network formation dynamics can only reveal properties of utility functions up to choice-equivalence. Therefore, without loss of generality, I will assume that $V_i(\varnothing) = 0$ and, whenever utilities are conservative, the corresponding potential satisfies $\Phi(\varnothing) = 0$. In order to study the structure of this equivalence classes of utilities, it is useful to analyze a case in which conservativeness fails.

\begin{example}
    Let $d^{\textrm{in}}_i(g)$ and $d^{\textrm{out}}_i(g)$ be the in and out-degrees of individual $i$ in network $g$. Suppose that utility functions are given by
    \begin{align}
        V_i(g) = d^{\textrm{in}}_i(g) d^{\textrm{out}}_i(g).
    \end{align}
    Note that this implies that the marginal utility that agent $i$ receives from creating a link is $d^{\textrm{in}}_i(g)$. These utility functions do not satisfy conservativeness. In order to see this, consider the empty network $g = \varnothing$ and three different agents $i$, $j$ and $k$. To evaluate conservativeness, we need to evaluate whether Equation \eqref{eq:conservative_utility} holds for all links and all networks. Consider first adding the link $ij$ and then the link $jk$. The sum of the corresponding marginal utilities is
    \begin{align}
        d^{\textrm{in}}_i(\varnothing) + d^{\textrm{in}}_j(\{ij\}) = 1.
    \end{align}
    Adding the links in the reverse order yields
    \begin{align}
        d^{\textrm{in}}_j(\varnothing) + d^{\textrm{in}}_i(\{jk\}) = 0.
    \end{align}
    Therefore, the conservativeness condition is violated for these changes.
\end{example}

In the example above, we see that the conservativeness condition is violated because the marginal utility of adding link $jk$ for agent $j$ is affected by agent $i$'s action, but agent $j$'s action of adding $jk$ does not affect agent $i$'s marginal utility symmetrically. Intuitively, this means that for conservativeness to hold, agents' effects on each other's marginal utilities should be symmetric. Does this mean that agents' utilities cannot depend on others' actions if they are conservative? Not necessarily. For example, utilities could depend on actions that do not affect the link formation incentives of the player. Let $g^{-i} \equiv \{ i'j' \in g : i' \ne i \}$ be the subset of the network which $i$ does not participate in. If the utility functions $V_i(g)$ are conservative, then the functions $V_i(g) + f_i(g^{-i})$ will also be conservative for arbitrary functions $f_i$. This shows that conservativeness can be preserved by adding structure that does not affect the link formation incentives. However, we can make a stronger statement that solidifies the intuition that the effect of players' actions on each other's marginal utilities must be symmetric in order for utilities to be conservative.

\begin{proposition} \label{prop:value_sharing}
    For any set of conservative utility functions $V$, there exists a unique set of parameters $(\alpha(g))_{g \in {\cal G}} \in \mathbb{R}^{{\cal G}}$ such that the utilities
    \begin{align}
        \Tilde{V}_i(g) \equiv \sum_{g_0 \subseteq g} \alpha(g_0) \mathbbm{1}\{ \exists j : ij \in g_0 \}
    \end{align}
    are choice-equivalent to $V$. Particularly, the potential associated to $V$ will also be the potential associated to $\Tilde{V}$, and it satisfies
    \begin{align}
        \Phi(g) = \sum_{g_0 \subseteq g} \alpha(g_0).
    \end{align}
    Furthermore, $\Tilde{V}$ is the unique set of utility functions that are choice-equivalent to $V$ and satisfy $\Tilde{V}_i(g^{-i}) = 0$ for all agents $i$ and networks $g$.
\end{proposition}


Proposition \ref{prop:value_sharing} shows that we can characterize the equivalence classes of conservative utility functions in an alternative way to the potential. By removing the part of the utility function that depends solely on others' actions, by setting $\Tilde{V}_i(g) = V_i(g) - V_i(g^{-i})$, we obtain a representation that has an intuitive interpretation. A network $g_0$ has some value $\alpha(g_0)$, which is received by agent $i$ under network $g$ if two conditions are satisfied: $g_0$ is a subgraph of $g$, and agent $i$ participates in the structure by having at least one outward-going link in $g_0$. Therefore, removing the part of $V$ that is orthogonal to other agents' actions induces a ``value-sharing'' utility structure where link-switching actions affect all participants of the subgraphs $g_0$ equally. This clearly reveals the connection between conservativeness and the symmetry of the effect of agents' actions on each others' utilities.

\begin{example} \label{ex:degree_potential}
    Suppose the utility of agent $i$ is proportional to their out-degree:
    \begin{align}
        V_i(g) = \alpha_0 d_i^{\textrm{out}}(g).
    \end{align}
    These utilities are conservative by Proposition \ref{prop:conv}. Note that this corresponds to a value-sharing utility where the groups are single dyads. That is, the values $\alpha$ in Proposition \ref{prop:value_sharing} are
    \begin{align}
        \alpha(g) =
        \begin{cases}
            \alpha_0 & \textrm{if } |g| = 1, \\
            0 & \textrm{otherwise}.
        \end{cases}
    \end{align}
    The potential is, therefore,
    \begin{align}
        \Phi(g) = \sum_{ij \in g} \alpha_0 = \alpha_0 \sum_{i \in {\cal N}} d_i^{\textrm{out}}(g).
    \end{align}
    This yields a stationary distribution that is an ERGM where the sufficient statistic is the total number of links in the network\footnote{This is the model considered in \citet{Mele2017AFormation}, with the additional simplification of constant values. This paper also considers groups of the form $\{ij, ji\}$ and $\{ij, jk, ki\}$.}.
\end{example}




\section{Forward-Looking Network Formation} \label{sec:complex_models}
Section \ref{sec:model} analyzed the case of agents that made choices based only on utilities that depend on the state of the network, independent of their beliefs of the future. In this section I extend this framework to allow individuals to incorporate their beliefs on others' future actions into their strategies.

In this new setup, agent $i$ receives a flow utility $v_i(g)$ is the state of the network is $g$. Agents discount the future at a rate $\rho$. Let $G_t$ be the random variable associated to the state of the network at time $t$. The meeting process is the same as in Section \ref{sec:model}, and the possible actions of agents are also to change the state of the link or to abstain. Agents still receive idiosyncratic shocks to their utility every time they meet another agent. Formally, the discounted utility an agent receives for a given realization of networks $(G_t)_{t \in \mathbb{R}_+}$ and meeting times $(\tau_n)_{n \in \mathbb{N}}$, with their associated idiosyncratic shocks $(\varepsilon_n)_{n \in \mathbb{N}}$\footnote{Note that, to simplify notation, the choices of switching and abstaining have been absorbed into $G_t$, such that shocks no longer need to be labeled as $\varepsilon_{ij}^k$.}, is
\begin{align} \label{eq:realization_utility}
    U[(G_t)_{t \in \mathbb{R}_+}, (\tau_n)_{n \in \mathbb{N}}, (\varepsilon_n)_{n \in \mathbb{N}}] = \rho \int_0^\infty e^{-\rho t} v_i(G_t) \, dt + \sum_{n=0}^\infty e^{-\rho \tau_n} \varepsilon_{n},
\end{align}
where the factor $\rho$ in front of the integral is introduced in order to interpret this discounted value as an average of the flow utilities. This expression for the discounted utility showcases that agents' actions affect their payoffs in two ways: they affect the future probabilities of network realizations, and they can potentially affect the distribution of times at which they make a choice, such that they can derive more value from being subject to more shocks. Since I want to focus on the economic forces arising from the dynamics of the network formation process, I will assume that agents are myopic to future shocks, and that they only incorporate the discounted future flow utility into their decisions.

I will focus on Markov-Perfect Equilibria of the network formation game, where the strategies of agents depend only on the state of the network. That is, at time $t$, an individual's strategy is measurable with respect to $G_t$. To this end, it is useful to define the following value function for agent $i$:
\begin{align} \label{eq:discounted_value}
    V_i(g) = \mathbb{E} \left[ \rho \int_0^\infty e^{-\rho t} v_i(G_t) \, dt \, \middle| \, G_0 = g \right],
\end{align}
where the expectation is taken with respect to realizations of future shocks and the beliefs of others' strategies. For a given value function, the decision process when agent $i$ meets agent $j$ is the same as in Equation \eqref{eq:disc_choice}, where agent $i$ chooses the action that maximizes expected future payoff plus the current shock. Thus, in an MPE, the value functions are given by Equation \eqref{eq:discounted_value}, where the network state evolves following the switching probabilities
\begin{align}
    p_{ij}(g) = [1+ \exp(-[V_{i}(\sigma_{ij}(g)) - V_{i}(g)])]^{-1}.
\end{align}
This means that an MPE is characterized by a fixed-point problem on the value functions $V$.

To more easily characterize the properties of Markov-Perfect Equilibria, it is useful to note that the value functions satisfy the following Hamilton-Jacobi-Bellman equation:
\begin{align}
    \rho V_i(g) = \rho v_i(g) + \sum_{i'j' \in {\cal D}} \lambda_{i'j'}(g) p_{i'j'}(g) [V_i(\sigma_{i'j'}(g)) - V_i(g)].
\end{align}
We can see that the inter-dependence of the present values of different individuals makes this problem significantly more complex than in the myopic case. To build some intuition, Proposition \ref{prop:MPE} outlines the limiting behavior of Markov-Perfect Equilibria. 

\begin{proposition} \label{prop:MPE}
For every $\rho \in (0,1)$ and every set of flow-utilities $v:{\cal N} \times {\cal G} \to \mathbb{R}$, a MPE exists. Additionally, let $(\rho_n)_{n \in \mathbb{N}}$ be a sequence of discount factors and let $V^n$ be utility functions in a MPE consistent with $\rho_n$. Then the following hold:
\begin{enumerate}
    \item If $\lim_{n \to \infty} \rho_n = 0$, then for every $i \in {\cal N}$ and all $g \in {\cal G}$,
    \begin{align}
        \lim_{n \to \infty} V^n_i(g) = \frac{1}{|{\cal G}|} \sum_{g' \in {\cal G}} v_i(g').
    \end{align}
    \item If $\lim_{n \to \infty} \rho_n = \infty$, then for every $i \in {\cal N}$ and all $g \in {\cal G}$,
    \begin{align}
        \lim_{n \to \infty} V^n_i(g) = v_i(g).
    \end{align}
\end{enumerate}
\end{proposition}

Proposition \ref{prop:MPE} implies that, in equilibrium, if individuals are infinitely impatient they are no different than in the myopic case. If they are infinitely patient, then only the long-term behavior of the network formation process is relevant. As such, the initial state of the process does not matter, as is evidenced in the limiting behavior of the value function.

Note that the results from Section \ref{sec:model} can be applied to an equilibrium if the myopic utilities are replaced by the present values. However, establishing a result like Proposition \ref{prop:conv}, where the stationary distribution of the network formation process is written in terms of model primitives, proves to be much harder. This is because the conservativeness condition on the value functions is a condition on fixed points of complicated functions of the flow utilities.

Despite the difficulty of establishing a condition like conservativeness, we can also ask what are the model primitives that yield a given Gibbs measure with a potential $\Phi$. This would provide a dynamic forward-looking microfoundation for ERGMs. Lemma \ref{lem:forward_looking} establishes that for an arbitrary potential $\Phi$ we can, indeed, construct model primitives that result in conservative utility functions with associated potential $\Phi$.

\begin{lemma} \label{lem:forward_looking}
For a given potential $\Phi$, define the value functions
\begin{align}
    V^\Phi_i(g) \equiv \Phi(g) - \Phi(g_{-i}).
\end{align}
Then for the flow utilities
\begin{align}
    v^\Phi_i(g) \equiv V^\Phi_i(g) - \frac{1}{\rho} \sum_{i'j' \in {\cal D}} \lambda_{i'j'}(g) p_{i'j'}^\Phi(g) [V_i^\Phi(\sigma_{i'j'}(g)) - V_i^\Phi(g)],
\end{align}
where $p^\Phi$ are the switching probabilities induced by $V^\Phi$, there exists a MPE where the resulting value functions are $V^\Phi$, and the corresponding potential is $\Phi$.
\end{lemma}

Lemma \ref{lem:forward_looking} implies that any Gibbs measure over the space of networks can be microfounded with a forward-looking model, in addition to the myopic microfoundation in Section \ref{sec:model}. However, with these results, we cannot guarantee that the MPE for which the utility functions have an associated potential $\Phi$ is the unique MPE for the flow utilities $v^\Phi$.

\begin{example} \label{ex:forward_looking_degree_potential}
    Suppose we want to construct a forward-looking model whose stationary distribution is an ERGM with potential
    \begin{align}
        \Phi(g) = \alpha_0 \sum_{i \in {\cal N}} d^{\textrm{out}}_i(g),
    \end{align}
    as in Example \ref{ex:degree_potential}. Additionally, suppose that $\lambda_{ij}(g) = \lambda$ for all $ij$ and $g$. From Example \ref{ex:degree_potential}, we know that
    \begin{align}
        V^\Phi_i(g) = \alpha_0 d^{\textrm{out}}_i(g).
    \end{align}
    Since the only switches $\sigma_{i'j'}$ that have a non-zero marginal value are those for which $i' = i$, we can easily calculate $v^\Phi$. The ERGM distribution, then, arises from an MPE associated to the following flow utilities:
    \begin{align}
        v_i^\Phi(g) = \alpha_0 d^{\textrm{out}}_i(g) - \frac{\lambda}{\rho} \left[ \frac{N - d^{\textrm{out}}_i(g) - 1}{1 + e^{\alpha_0}} - \frac{d^{\textrm{out}}_i(g)}{1 + e^{-\alpha_0}} \right].
    \end{align}
    Intuitively, this means that, relative to the myopic case, we need to increase the marginal value of each link in the flow utilities. Specifically, the marginal value of each additional link must be shifted up to
    \begin{align}
        \alpha_0' = \alpha_0 + \frac{\lambda}{\rho} \left[ \frac{1}{1+e^{\alpha_0}} + \frac{1}{1+e^{-\alpha_0}} \right].
    \end{align}
\end{example}

\section{Large Dense Networks} \label{sec:large_networks}
The results in the previous section gave conditions under which our network formation process can be characterized in terms of a potential, and established some properties of the structure of such processes. This allowed us to identify the most probable networks with Nash equilibria of the deterministic game. In this section, I study processes generated by a particular class of conservative utility functions as $N$ grows large, revealing interesting typical behaviors that arise form the interplay of incentives and the exponentially increasing size of network space.

In order to make the number of players grow large in a consistent way, I consider a set of conservative utilities parametrized by a finite set of $L$ parameters $\Omega \in \mathbb{R}^L$. For a given $N$, let $V^N(\Omega)$ be the value-sharing representation of the utilities, $\Phi^N(\Omega)$ be the potential and $\pi^N(\Omega)$ be the stationary probabilities. Similarly, I index the corresponding set of agents, networks and dyads by ${\cal N}^N$, ${\cal G}^N$ and ${\cal D}^N$, respectively.

As mentioned before, the largest difficulty in analyzing the behavior of this model is characterizing the denominator in Equation \eqref{eq:eq_dist}, which is intractable. This is because the number of terms in this sum is $2^{N(N-1)}$. Despite the difficulty in calculating this sum, this object encodes useful information about the system statistics. Following the physics literature \citep{huang2008statistical}, I now define the \textit{scaled partition function} (or scaled normalization factor).

\begin{definition}
    The scaled partition function of the system for a set of parameters $\Omega$ is
    \begin{align}
        \zeta^N(\Omega) = \frac{1}{N^2} \log\left( \sum_{g \in {\cal G}^N} \exp(\Phi^N(g \,|\, \Omega)) \right).
    \end{align}
\end{definition}

To understand the $1/N^2$ re-scaling intuitively, note that to obtain dense networks (networks where agents typically form a non-vanishing fraction of their connections), the marginal utilities of changing links must be of order $N$. Since the potential is capturing the incentives of $N$ agents, it will typically be of order $N^2$. Given this scaling of the potential, together with the fact that the size of network space scales as $2^{N^2}$, we would expect $\zeta^N(\Omega)$ to be of order $1$.

\begin{example} \label{ex:degree_zeta}
    In the particular case of ERGMs, the partition function encodes information about the sufficient statistics of the system. Letting $\zeta^N_{\textrm{ERGM}}(\boldsymbol{\beta})$ be the partition function for the ERGM distribution in Equation \eqref{eq:ERGM}, where the set of parameters is $\Omega = \boldsymbol{\beta}$. Differentiating with respect to $\beta_i$ yields
    \begin{align}
        \frac{\partial \zeta^N_{\textrm{ERGM}}(\boldsymbol{\beta})}{\partial \beta_i} = \frac{1}{N^2} \sum_{g \in {\cal G}^N} \pi^N_{\textrm{ERGM}}(g \,|\, \boldsymbol{\beta}) \left( \frac{\partial \Phi^N_{\textrm{ERGM}}(g \,|\, \boldsymbol{\beta})}{\partial \beta_i} \right) \equiv \frac{1}{N^2} \mathbb{E}_{\pi^N_{\textrm{ERGM}}} \left[ S_i(g) \right],
    \end{align}
    where $\mathbb{E}_{\pi^N_{\textrm{ERGM}}}$ denotes the expectation over ${\cal G}^N$ under the ERGM distribution. 

    As a concrete example, consider the utility function in Example \ref{ex:degree_potential}:
    \begin{align}
        V_i^N(g \,|\, \Omega) = \alpha_0 d^{\textrm{out}}_i(g),
    \end{align}
    such that $\Omega = \{\alpha_0\}$. Then differentiating the partition function with respect to $\alpha_0$ yields
    \begin{align}
        \frac{\partial \zeta^N(\Omega)}{\partial \alpha_0} = \frac{1}{N^2} \mathbb{E}_{\pi^N} \left[ \sum_{i=1}^N d^{\textrm{out}}_i \right],
    \end{align}
    which corresponds to the average density of links in the network under the corresponding Gibbs measure.
\end{example}

This result also has an interpretation in the dynamic model: derivatives of the partition function with respect to model parameters yield the long-time averages of the sufficient statistics of an ERGM. As illustrated in Example \ref{ex:degree_zeta}, the partition function can also contain information about the utility functions of agents.

I will now study specific forms of conservative utilities that generate dense networks as $N$ grows large. An interesting aspect of the model that arises from this analysis is that the scaled partition function of the system will not only be useful for calculating statistics of the limiting distribution, but rather will be the central object that characterizes the typical behavior of the system as it navigates the space of networks.

\subsection{Network motifs}
As illustrated by Proposition \ref{prop:value_sharing}, conservativeness of utilities is tightly related to the value of social structures. We can study the large $N$ behavior of the model if the value assigned to structures depends only on the topology of the structure. Specifically, I will study the case where there is a fixed set of structures which agents value, and they derive value proportional to the number of such structures present in the network which they are a member of. Following the graph theory literature, I call these structures \textit{motifs}. This characterization is particularly amenable to our analysis, since it allows a clean connection between the large $N$ asymptotics of the networks and the model primitives.

Formally, a motif $m$ is a network over a set ${\cal N}_m = \{1, \ldots, n_m\}$ of $n_m$ nodes that has $e_m \equiv |m|$ edges. The number of realizations of a motif $m$ in network $g$ is defined as the number of injective maps $\varphi: {\cal N}_m \to {\cal N}^N$ such that the subgraph obtained from applying this mapping to the nodes and edges of $m$ is present in $g$. This quantity is given by 
\begin{align}
    T^N(g) \equiv \sum_{\substack{\varphi: {\cal N}_m \to {\cal N}^N \\ \varphi \textrm{ injective}}} \Bigg( \underbrace{ \prod_{jk \in m} \mathbbm{1}\{\varphi(j)\varphi(k) \in g\}}_{\textrm{motif is present in network}} \Bigg).
\end{align}
Additionally, we can define the number of instances of motif $m$ that agent $i$ belongs to in network $g$ as 
\begin{align}
    T^N_i(m,g) \equiv \sum_{\substack{\varphi: {\cal N}_m \to {\cal N}^N \\ \varphi \textrm{ injective}}} \Bigg( \underbrace{ \prod_{jk \in m} \mathbbm{1}\{\varphi(j)\varphi(k) \in g\}}_{\textrm{motif is present in network}} \Bigg) \underbrace{\mathbbm{1}\{\exists j,k \in {\cal N}_m: \varphi(j) = i, jk \in m\}}_{i \textrm{ participates in the subnetwork}}.
\end{align}

\begin{figure}
    \begin{center}
    \caption{Motif counting}
    \includegraphics[width = 0.7 \linewidth]{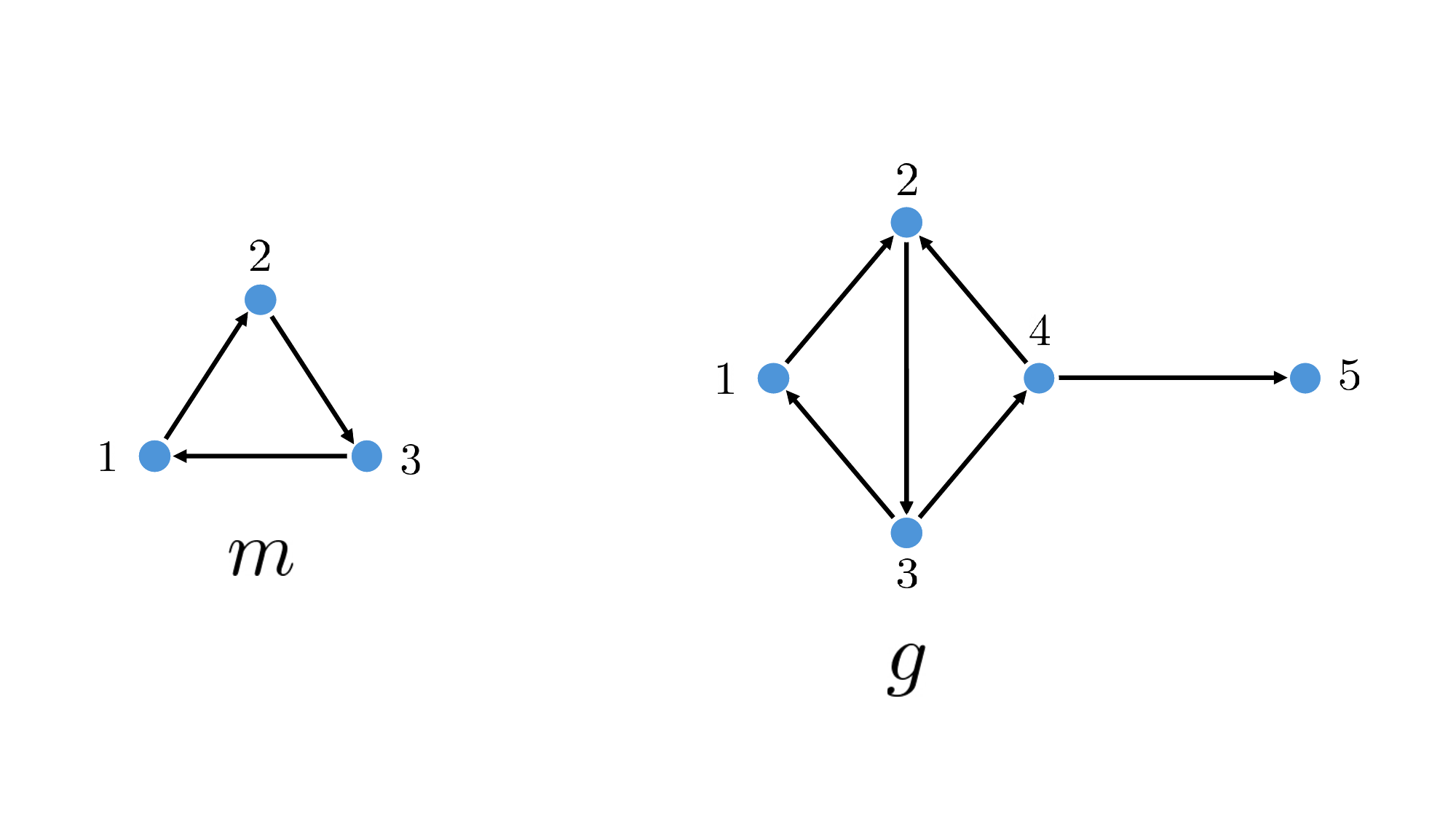}
    \label{fig:motif_counting}
    \end{center}
    \footnotesize{\textit{Notes.} This figure illustrates the process of counting the realizations of motif $m$ in network $g$. There are $T^5(m,g) = 6$ injective mappings from the nodes of $m$ to the nodes of $g$ that preserve structure. However, these correspond to only 2 distinct realizations of the motif: the subgraphs corresponding to the nodes $\{1,2,3\}$ and $\{2,3,4\}$. To avoid overcounting, we can divide the number of injective mappings by $h_m = 3$, the number of structure-preserving mappings from the nodes of $m$ to themselves. The density of motif $m$ in network $g$ is $b^5(m,g) = T^5(m,g)/5^3 = 0.048.$}
\end{figure}

I will study the case where agents derive utility $a^N_m$ from being in a subnetwork $g_0$ if $g_0$ is an instance of motif $m$. That is, a structure $g_0$ has value $a^N_m$ if there exists an injective map $\varphi$ such that $g_0$ is the result of applying $\varphi$ to the nodes of $m$, and 0 otherwise. The total utility of agent $i$ is, then,
\begin{align}
    V_{i,m}^N(g) = \frac{a_m^N}{h_m} T^N_i(m,g),
\end{align}
where $h_m$ is the number of injective maps $\varphi':{\cal N}_m \to {\cal N}_m$ such that $ij \in m \iff \varphi'(i) \varphi'(j) \in m$. This accounts for the possible overcounting when summing over all injective maps, as illustrated in Figure \ref{fig:motif_counting}. Under these utilities, the corresponding potential is
\begin{align}
    \Phi_{m}^N(g) = \frac{a_m^N}{h_m} T^N(m,g),
\end{align}
which follows from the characterization in Proposition \ref{prop:value_sharing}.

The scaling of the number of injective maps from ${\cal N}_m$ to ${\cal N}^N$ depends on $n_m$, and it determines how the utilities of agents scale as $N$ grows large. There are $N^{n_m}$ maps from ${\cal N}_m$ to ${\cal N}^N$, and a vanishing fraction of these will be non-injective as $N$ grows large, so it is common to define the density of motif $m$ in network $g$ as\footnote{In the graph theory literature, this is referred to as the \textit{subgraph density}. If the maps used to define $T^N$ are not required to be injective, it is instead called the \textit{homomorphism density}.}
\begin{align}
    b^N(m,g) \equiv \frac{T^N(m,g)}{N^{n_m}}.
\end{align}
Similarly, agent $i$'s participation density in motif $m$ is
\begin{align}
    b_{i}^N(m,g) \equiv \frac{T^N_i(m,g)}{N^{n_m}}.
\end{align}
Intuitively, if the utilities of agents are such that the potential does not scale like $N^2$, the density of the network will typically be 0 or 1. To study non-trivial dense networks, I will study utilities that yield potentials that scale proportional to $N^2$. Note that $b^N(m,g)$ ranges from 0 to 1 over the space of networks, so we can achieve the correct scaling by studying values $a_m^N$ that have the following form:
\begin{align}
    a^N_m = \frac{N^2}{N^{n_m}} h_m a_m.
\end{align}
The utility of agent $i$, then, can be written in terms of the participation density of $i$ in $m$:
\begin{align}
    V_{i,m}^N(g) = N^2 a_m b^N_i(m,g).
\end{align}
If we now consider that agents derive utility from a set of motifs ${\cal M}$, the total utility of agent $i$ is
\begin{align} \label{eq:motif_utility}
    V^N_{i,{\cal M}}(g \,| \, {\bf a}) = N^2 \sum_{m \in {\cal M}} a_m b_{i}^N(m,g),
\end{align}
where ${\bf a} \in \mathbb{R}^{{\cal M}}$. The potential corresponding to these utilties is
\begin{align}
    \Phi^N(g \,| \, {\bf a}) = N^2 \sum_{m \in {\cal M}} a_m b^N(m,g).
\end{align}
This characterization of the utility functions allows us to tractably analyze the behavior of the system as $N \to \infty$.

\begin{proposition} \label{prop:motif_partition}
Suppose $a_m > 0$ for all motifs $m$ with $e_m > 1$. Then the limiting partition function is given by
\begin{align} \label{eq:motif_partition}
    \zeta({\bf a}) = \max_{\rho \in [0,1]} \Bigg[ \underbrace{\sum_{m \in {\cal M}} a_m \rho^{e_m} }_{\textrm{motif values}} + \underbrace{H(\rho)}_{\textrm{entropy}} \Bigg].
\end{align}
where $H(p) \equiv -p \log(p) - (1-p) \log(1-p)$ is the entropy of a Bernoulli random variable with parameter $p$. Additionally, if the maximizer $\rho^*({\bf a})$ is unique, then as $N \to \infty$ the networks generated by the model become indistinguishable from those generated by an Erd\"os-R\'enyi random graph model with parameter $\rho^*({\bf a})$\footnote{See Appendix \ref{sec:app_graph_limits} for a detailed treatment of graph limits}.
\end{proposition}

Proposition \ref{prop:motif_partition} is similar to Theorem 2 in \citet{Mele2017AFormation}, and builds on it by extending the result to a microfounded model with arbitrary structures using the characterization in Proposition \ref{prop:value_sharing}. This result essentially reduces the dimensionality of the complex network formation process to a single one-dimensional optimization process, where the maximizer $\rho^*$ can be interpreted as the typical density of the resulting networks. The nature of this optimization process is interesting, since no single agent is solving the problem in Equation \eqref{eq:motif_partition}. To understand it, it is useful to analyze where the entropy term comes from. This term arises from the fact that the number of networks over $N$ nodes with density $\rho$, to leading exponential order, is approximately $2^{N^2 H(\rho)}$. Therefore, as agents stochastically try to optimize their utility by building motifs, they get ``stuck'' exploring an exponentially growing region of network space, leading to a trade-off between utility and entropy.

It is important to note is that this analysis is restricted to positive motif values for motifs that involve more than one link, since the characterization in terms of typical densities can break down when these values are negative, as shown in \citet{Mele2017AFormation}. This is of particular interest when analyzing the problem of estimation of model parameters, but I will restrict my attention to positive values, since this is still a rich enough case where there are non-trivial economic forces at play. One particularly interesting phenomenon that arises in the large $N$ limit is the emergence of phase transitions.

\subsection{Phase transitions} \label{subsec:phase_transitions}
An astonishing result in the theory of graph limits is the possibility of discontinuous changes in the aggregate properties of the model caused by continuous changes in parameters. These type of phenomena fall under the category of \textit{phase transitions} in the physics and applied mathematics literature, and the properties of systems near critical points (the parameter values where phase transitions occur) have been studied extensively\footnote{For an detailed overview of phase transitions in physics, see \citet{goldenfeld2018lectures}. For an application of the theory of phase transitions to ERGMs, see \citet{aristoff2017phase}.}. In the case of our model, we will see that varying the values of motifs can induce a phase transition, causing a discontinuous change in the typical density of the networks. In order to build some intuition, it is useful to map the maximization problem in Proposition \ref{prop:motif_partition} into a fixed-point problem.

\begin{lemma}\label{lem:fixed_point}
    Define the marginal value of motifs with respect to the density as
    \begin{align} \label{eq:K_rho}
        K_{\cal M}(\rho) \equiv \sum_{m \in {\cal M}} a_m e_m \rho^{e_m - 1}.
    \end{align}
    Then the maximizer $\rho^*$ in Proposition \ref{prop:motif_partition} is a fixed point of the function
    \begin{align} \label{eq:motif_fixed_point}
        R(\rho) \equiv \left[ 1 + \exp(-K_{\cal M}(\rho)) \right]^{-1}.
    \end{align}
\end{lemma}

\begin{figure}
    \begin{center}
    \caption{Phase transition induced by motif values}%
    \subfloat[Behavior of the fixed points of $R(\rho)$]{{\includegraphics[width=0.49\linewidth]{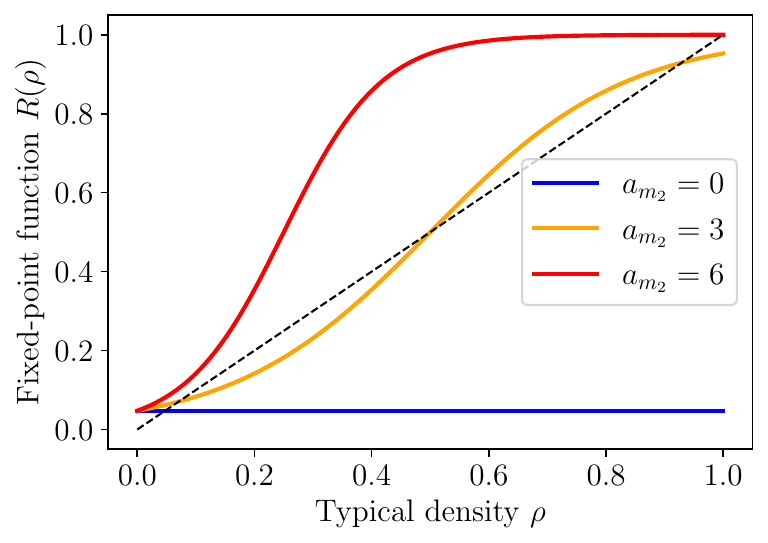} }}%
    \subfloat[Density phase diagram]{{\includegraphics[width=0.49\linewidth]{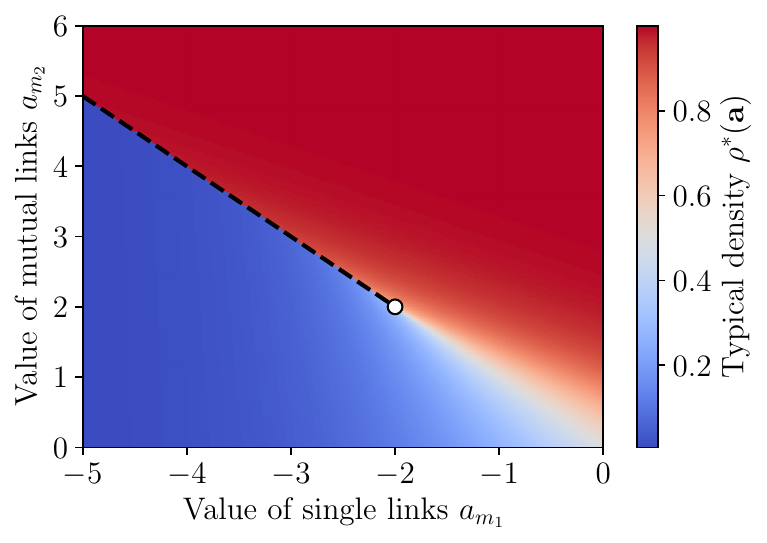} }}
    \label{fig:density_phase_transition}
    \end{center}
    \footnotesize{\textit{Notes.} This figure illustrates the mechanisms behind phase transitions with two motif values. Panel (a) shows the behavior of the fixed points of $R(\rho)$ as the value of mutual links $a_{m_2}$ is varied, while the value of single connections is kept constant at $a_{m_1} = -3$. Panel (b) shows the value of the typical density for different values of the motifs $m_1$ and $m_2$.}
\end{figure}

Lemma \ref{lem:fixed_point} allows for a graphical representation of the concentration result in Proposition \ref{prop:motif_partition}. To visualize this, let us consider the case of two motifs: ${\cal M} = \{m_1, m_2\}$, where $m_1 \equiv \{12\}$ corresponds to single links and $m_2 \equiv \{12, 21\}$ corresponds to mutual links between two individuals. The function $R(\rho)$ is shown in panel (a) of Figure \ref{fig:density_phase_transition}. We can see that, for a fixed value of motif $m_1$, low and high values of $a_{m_2}$ produce unique fixed points, but for intermediate values there can be a bifurcation, and a new pair of fixed points can emerge. In this specific example, for $a_{m_2}=0$ there is only a low density fixed point. Then, as $a_{m_2}$ increases, a pair of high density fixed points appear, and as it is increased even more, the low density fixed point gets annihilated altogether. Since the maximizer must always be a fixed point, this means that at some point the system must discontinuously change from the low density fixed point to the high density one. As shown in panel (b) of Figure \ref{fig:density_phase_transition}, this discontinuous jump occurs for a wide range of values of $a_{m_1}$, and it defines a clear change between a phase of low-density and high-density networks. Thus, reducing the value of mutual links across the line of phase transitions can lead to the discontinuous collapse of the networks in the model.

In order to have a clear intuition of the economic forces that are causing this discontinuous transition, it is useful to relate this result to the case where there are no idiosyncratic shocks in the choices of agents. As shown in \citet{Mele2017AFormation}, this model converges to a local maximum of the potential, which corresponds to a Nash equilibrium of the game. Analyzing the potential, then, provides a convenient way to compute the Nash equilibria of the game, which are a natural point of comparison with our previous results. 

\begin{figure}
    \begin{center}
    \caption{Phase transition in the 5-star motif model}
    \includegraphics[width = 0.6 \linewidth]{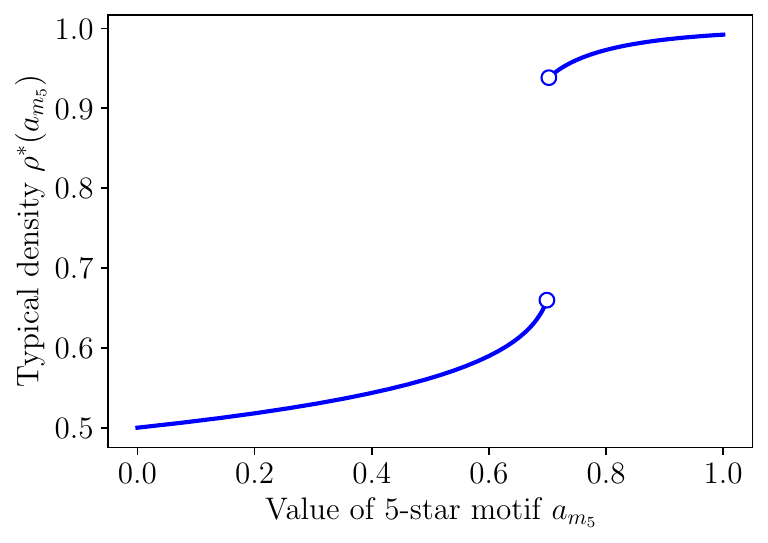}
    \label{fig:star_phase_transition}
    \end{center}
    \footnotesize{\textit{Notes.} This figure shows the typical density of the network under the 5-star motif utility when the maximizer in Proposition \ref{prop:motif_partition} is unique.}
\end{figure}

\begin{lemma}\label{lem:nash_equilibria}
    Consider the model with motifs ${\cal M} = \{m_1, m_2\}$. The Nash equilibria of the deterministic game satisfy:
    \begin{itemize}
        \item if $a_{m_1} < - a_{m_2}$, there is a unique Nash equilibrium given by the empty network.
        \item if $a_{m_1} > - a_{m_2}$, all networks $g$ such that $ij \in g \iff ji \in g$ are Nash equilibria.
    \end{itemize}
\end{lemma}
The result in Lemma \ref{lem:nash_equilibria} shows that there is a phase transition in this model even in the absence of the stochasticity in agents' decisions, since the density of the equilibrium network jumps from 0 to 1 as the line $a_{m_1} = - a_{m_2}$ is crossed\footnote{This is related to the concept of zero-temperature phase transitions in statistical physics. See \citet{goldenfeld2018lectures} for a detailed discussion.}. Comparing this to Figure \ref{fig:density_phase_transition}, we see that this is precisely the line that defines the phase transition below some threshold $a_{m_1}^*$. Therefore, the frictions from the meeting process can \textit{remove} the phase transition if the incentives are not strong enough. However, this is not true in general. Consider the case where there is a single motif: ${\cal M} = \{s_5\}$, where $s_5 \equiv \{12,13,14,15,16\}$ is a directed star network with 1 central node and 5 peripheral nodes. In this case, for all $a_{s_5} > 0$, the unique Nash equilibrium is the complete network. As shown in Figure \ref{fig:star_phase_transition}, the typical density exhibits a phase transition in this model as well. This shows that there is a non-trivial interplay between the incentives of players and the frictions in exploring network space for phase transitions to emerge. Therefore, we can obtain phase transitions that are incentive-driven or entropy-driven.

An important ingredient for phase transitions to appear is that agents must derive utility from structures that are ``sufficiently complex'' to generate these effects. For example, if the only motif we considered were a single link, the typical density would be a continuous function of the utility of this motif. A similar complexity requirement was found in the analysis of supply networks in \citet{elliott2022supply}, where there is a discontinuity in the probability of successful production as a function of the probability that individual production ties are successful. They find that the complexity of production, defined as the number of different inputs required to produce some good, is fundamental in generating this discontinuity. It seems, therefore, that the feedback generated by having multiple relationships is a recurring feature of models that generate these types of phenomena.

\subsection{Heterogeneous agents}
Our previous analysis showed that the network formation model can generate non-trivial phenomena even when all agents have the same utility. In this section, I expand on this by analyzing the network formation game for heterogeneous agents. Heterogeneity is a natural property of real network formation processes. For example, individuals have the tendency to form disproportionately more connections with people similar to themselves than to others. This is a phenomenon known as homophily \citep{McPherson2001BirdsNetworks, Pin2016StochasticHomophily}. Another setting where heterogeneity is important is when networks are spatially embedded, such as in trade routes, since distance between agents becomes an important factor when determining whether or not to form relationships. Here, I study how heterogeneity in preferences interacts with network effects to determine the aggregate properties of large networks.

To characterize agents' heterogeneity, I classify them into a finite number of types. Let $\Theta$ be the set of agents' types, with $|\Theta| = L < \infty$. For a given number of agents $N$, their types are given by $\theta_1^N, \ldots, \theta_N^N \in \Theta$. To tractably analyze the large $N$ behavior of the network formation process, I study a sequence of type vectors whose limiting distribution is known. Specifically, let $\hat{{\bf w}}^N \in \Delta(\Theta)$ be the empirical distribution of agents' types. That is,
\begin{align*}
    \hat{w}^N_\theta \equiv \frac{1}{N} \sum_{i=1}^N \mathbbm{1}\{\theta_i^N = \theta\} \quad \forall \theta \in \Theta.
\end{align*}
As $N \to \infty$, let $\hat{{\bf w}}^N \to {\bf w}$ for some ${\bf w} \in \Delta(\Theta)$ with full support over $\Theta$.

Agents still derive value from network motifs, which only depend on network structure and not on agents' types. In addition to this, I now assume agents' utilities depend on the distribution of types of their neighbors. This allows us to systematically include heterogeneity in the agents, while also maintaining group-formation incentives. Define the scaled empirical distribution of agent $i$'s neighbors in network $g$ as
\begin{align}
    \hat{z}_{i,\theta}^N(g) = \sum_{j=1}^N \mathbbm{1}\{\theta_j^N = \theta\} \mathbbm{1}\{ij \in g\} \quad \forall \theta \in \Theta.
\end{align}
Note that $\hat{{\bf z}}_i^N(g)$ need not be a probability distribution like $\hat{{\bf w}}^N$. Agents' utility functions now include a component $u: \Theta \times \mathbb{R}^\Theta_+ \to \mathbb{R}$, which depends on their own type and the scaled empirical distribution of their neighbors' types. With this, the utility of agent $i$ in network $g$ is determined by the motif incentives, given by Equation \eqref{eq:motif_utility}, and the value of their neighborhood:
\begin{align}
    V_i^N(g) = \underbrace{V^N_{i,{\cal M}}(g \,| \, {\bf a})}_{\textrm{motif values}} + \underbrace{u_{\theta_i^N}[\hat{{\bf z}}_i^N(g)]}_{\textrm{neighborhood value}}.
\end{align}
To be able to characterize the limiting distribution of networks, I make some regularity assumptions on $u$.
\begin{assumption} \label{A:neighborhood_utility}
    The neighborhood utility function $u: \Theta \times \mathbb{R}^\Theta_+ \to \mathbb{R}$ has the following properties:
    \begin{itemize}
        \item \textit{Homogeneity}: for all $\theta \in \Theta$, ${\bf z} \in \mathbb{R}^\Theta_+$, and $c > 0$, $u$ satisfies $u_\theta[c {\bf z}] = c u_\theta[{\bf z}]$.
        \item \textit{Boundedness for probability distributions}: there exists a $K > 0$ such that $|u_\theta[{\bf z}]| < K$ for all $\theta \in \Theta$ and ${\bf z} \in \Delta(\Theta)$.
        \item \textit{Concave decomposition}: for all $\theta \in \Theta$, the function $u_\theta$ can be decomposed into
        \begin{align}
            u_\theta({\bf z}) = \sum_{\theta' \in \Theta} c_{\theta \theta'} z_{\theta'} + \Tilde{u}_\theta({\bf z}),
        \end{align}
        where $\Tilde{u}_\theta$ is concave and monotone. That is, for all ${\bf z}, {\bf z}' \in \mathbb{R}^\Theta_+$, and $\delta \in(0,1)$, $\Tilde{u}_\theta$ satisfies $\tilde{u}_\theta[\delta {\bf z} + (1-\delta) {\bf z}'] \ge \delta \tilde{u}_\theta[{\bf z}] + (1-\delta) \tilde{u}_\theta[{\bf z}']$, and if $z_{\theta'} \le z'_{\theta'}$ for all $\theta'$, then $\tilde{u}_\theta({\bf z}) \le \tilde{u}_\theta({\bf z}')$.
    \end{itemize}
\end{assumption}

\begin{example}
An example of a neighborhood distribution utility is one that counts the number of agents that an individual is connected to:
\begin{align}
    u_\theta[{\bf z}] = \alpha_0 \sum_{\theta' \in \Theta} z_{\theta'}.
\end{align}
For a given network, then, this is
\begin{align}
    u_{\theta_i^N}[\hat{{\bf z}}_i^N(g)] = \alpha_0 d_i^{\textrm{out}}(g),
\end{align}
which is the utility function in the previous examples.
\end{example}

In the case of homogeneous agents, the distribution resulting from the network formation process concentrated around networks with some typical density. To extend this result to heterogeneous agents, we need to allow more freedom in the characterization of the limiting distribution. Particularly, it will be sufficient to characterize the typical density of links between types $\theta$ and $\theta'$. Formally, let ${\cal K}_\Theta$ be the set of functions $\psi: \Theta^2 \to [0,1]$, which I denote as \textit{kernels}. Intuitively, if the process converges to a density kernel $\psi^*$, the value $\psi^*(\theta,\theta')$ will correspond to the fraction of agents of type $\theta'$ that agents of type $\theta$ will typically have a connection with.

Another generalization that must be made to account for agent heterogeneity is of the method of calculating motif densities. Intuitively, the density of motif $m$ in an Erd\"os-R\'enyi network with density $\rho$ is $\rho^{e_m}$. This allows for a clean characterization of the limiting typical networks, as seen in Proposition \ref{prop:motif_partition}. In order to generalize this counting to a network with typical density kernel $\psi$, the density must now depend on the kernel and the number of agents of a given type. Formally, for a kernel $\psi \in {\cal K}_\Theta$ we define the density of motif $m$ as\footnote{For a formal justification for the introduction of this limiting object, see Appendix \ref{sec:app_graph_limits}.}
\begin{align}
    b[m,\psi;{\bf w}] \equiv \sum_{\boldsymbol{\theta} \in \Theta^{n_m}} \left( \prod_{i \in {\cal N}_m} w_{\theta_i} \right) \left( \prod_{ij \in m} \psi_{\theta_i \theta_j} \right).
\end{align}
Note that this definition of motif density is equal to $\rho^{e_m}$ for the kernel satisfying $\psi_{\theta \theta'} = \rho$ for all $\theta, \theta'$, which matches our intuition. With this definition, we can study how group formation incentives interact with neighborhood values.
 
\begin{theorem} \label{thm:mult_types_partition}
If $a_m > 0$ for all motifs with $e_m > 1$ and the neighborhood utility function satisfies Assumption \ref{A:neighborhood_utility}, then the limiting partition function is given by
\begin{align}
    \zeta({\bf a},u,{\bf w}) = \max_{\psi \in {\cal K}_\Theta} \Bigg[ \underbrace{\sum_{m \in {\cal M}}  a_m b[m,\psi; {\bf w}]}_{\textrm{motif values}} + \sum_{\theta \in \Theta} w_\theta \Bigg[ \underbrace{\sum_{\theta' \in \Theta} w_{\theta'} H(\psi_{\theta \theta'})}_{\textrm{entropy}} + \underbrace{u_\theta[(w_{\theta'} \psi_{\theta \theta'})_{\theta' \in \Theta}]}_{\textrm{neighborhood value}} \Bigg] \Bigg].
\end{align}
Furthermore, if the maximizer $\psi^*({\bf a},u,{\bf w})$ is unique, then as $N \to \infty$ the networks generated by the model become indistinguishable from those generated by a directed stochastic block model with edge probabilities between types $\theta$ and $\theta'$ given by the kernel $\psi^*({\bf a},u,{\bf w})(\theta,\theta')$.
\end{theorem}

Theorem \ref{thm:mult_types_partition} provides a similar reduction in dimensionality as Proposition \ref{prop:motif_partition}, where the complex model of network formation can be understood in terms of a limiting kernel. The same forces drive the result, in the sense that agents must explore an exponentially growing region of network space, and hence the appearance of an entropic term. However, now there is heterogeneity in agents' incentives, leading agents of different types to explore in different directions, which results in a potentially non-uniform kernel. In addition to the neighborhood incentives, there is still a global feedback in the form of the motif values. This model provides an additional dimension to the complex interplay between incentives and network formation frictions, in that network structure can now affect the composition of who agents interact with.
\section{A Simple Model of Trade} \label{sec:trade}
In order to illustrate the economic interpretation of the forces driving the phenomena introduced in Section \ref{sec:large_networks}, I study a simple model of trade. In this setting, assigning values to motifs and neighborhoods arises naturally from the incentives of firms.

\subsection{Technology and incentives}
\begin{figure}
    \begin{center}
    \caption{Distance in the trade model}
    \includegraphics[width = 0.5 \linewidth]{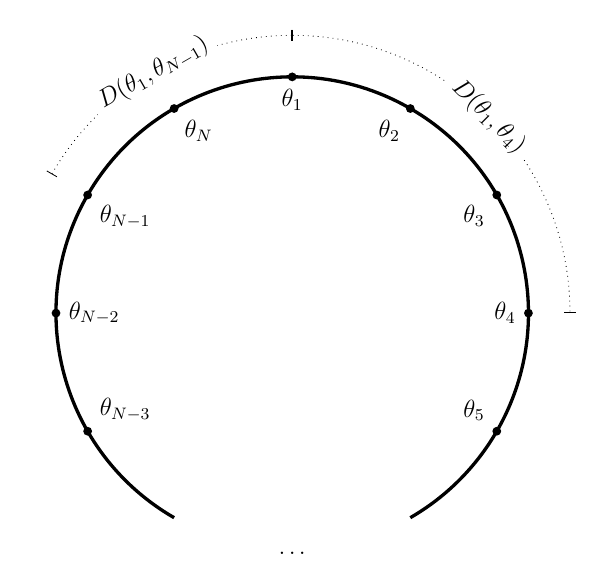}
    \label{fig:distance}
    \end{center}
    \footnotesize{\textit{Notes.} This figure illustrates the distance between firms of different types in the trade model model in Section \ref{sec:trade}. A specific realization of types for the case of $N$ firms is shown. The set of types is $\Theta = \{0,\frac{1}{L},\ldots,\frac{L-1}{L}\}$ and the distance is given by $D(\theta, \theta') = \min \{|\theta-\theta'|, 1-|\theta-\theta'|\}$. In this representation, the city has a circumference of 1.}
\end{figure}

Consider a set of firms that are spatially embedded in a circular city with unit circumference. Particularly, a firm's type is determined by its location, such that $\Theta = \{0,\frac{1}{L}, \ldots, \frac{L-1}{L}\}$. The main quantity that determines the incentives of a firm to trade with another is the distance to the other firm, defined as
\begin{align}
    D(\theta, \theta') = \min\{|\theta-\theta'|,1-|\theta-\theta'|\}.
\end{align}
This definition of distance is illustrated in Figure \ref{fig:distance}. As $N \to \infty$, the distribution of firms over $\Theta$ converges to ${\bf w}$.

Firms play a myopic network formation game, where the presence of a link $ij$ signifies that firm $i$ wishes to establish a trade route with firm $j$. To establish this link, firm $i$ faces a fixed cost of $\gamma D(\theta_i,\theta_j)$, which is proportional to the distance between the firms. If both firms wish to establish a trade route, this route becomes active and generates a revenue of $r$. Hence, for a fixed $N$, the utility of agent $i$ in network $g$ is
\begin{align}
    V^N_i(g \,|\, \gamma, r) = \sum_{j \in {\cal N}^N} [r \mathbbm{1}\{ji \in g\} - \gamma D(\theta_i, \theta_j) ] \mathbbm{1}\{ij \in g\}.
\end{align}
This utility can be decomposed into two parts: the value of participating in a mutual-links motif and the cost of establishing trade intentions with a neighborhood. 

In terms of the notation in section \ref{sec:large_networks}, firms assign value to a single motif, such that ${\cal M} = \{m_2\}$, where $m_2 = \{12,21\}$ is the ``mutual links'' motif. This motif has 2 permutations that leave it invariant, meaning that $h_{m_2} = 2$. Therefore, the proper scaling of the motif value is given by $a_{m_2} = r/2$. 

The cost of establishing trade intentions can be written as a neighborhood distribution utility, given by
\begin{align}
    u_\theta[{\bf z}] = - \gamma \sum_{\theta' \in \Theta} z_{\theta'} D(\theta,\theta').
\end{align}
Using this formulation, we can use our previous results on the limiting properties of the network formation game to study the effect of trading incentives on aggregate outcomes.

\subsection{Taxes on trade}
Characterizing the network formation game in terms of motif and neighborhood values allows us to formulate the problem of finding the typical density kernel $\psi^*$ that characterizes the limiting properties of the system. Following Theorem \ref{thm:mult_types_partition}, the limiting partition function of this model is given by
\begin{align} \label{eq:trade_partition}
    \zeta(\gamma, r) = \max_{\psi \in {\cal K}_\Theta} \left[ \sum_{\theta,\theta' \in \Theta^2} w_\theta w_{\theta'} \left( \frac{r}{2} \psi_{\theta \theta'} \psi_{\theta' \theta} + H(\psi_{\theta \theta'}) -\gamma \psi_{\theta \theta'} D(\theta,\theta') \right) \right],
\end{align}
and the typical density kernel $\psi^*$ is the maximizer. Interestingly, the solution to this problem can be characterized in terms of our solution to the problem with only motif values.

\begin{lemma} \label{lem:trade_fixed_point}
    Let $\rho^*({\bf a})$ be the typical density of the model with motifs ${\cal M} = \{m_1,m_2\}$, where $m_1 = \{12\}$ and $m_2 = \{12,21\}$, as in Section \ref{subsec:phase_transitions}. Suppose the maximizer $\psi^*$ in the problem above is unique. Then, point-wise, $\psi^*_{\theta \theta'}$ is given by $\rho^*({\bf a}(\theta,\theta'))$, where ${\bf a}(\theta,\theta') = (-\gamma D(\theta,\theta'),r/2)$.
\end{lemma}

Intuitively, this result says that the incentives in this model are such that there is effectively a network formation process with motif incentives \textit{for each pair} $(\theta,\theta')$, and these do not affect each other for different pairs. This means that our results for phase transitions in the homogeneous agent model would hold locally for this model. Therefore, for a fixed $\theta$, $\psi_{\theta \theta'}$ can have a discontinuity as $\theta'$ is varied. 

This characterization allows us to think about the effect of policy interventions on network structure. In particular, consider a tax rate $\tau$ on revenues generated from trade, such that the value of forming mutual links is now $(1-\tau) r$. Effectively, this intervention is equivalent to a decrease in the value of motif $m_2$, which can trigger a phase transition in the homogeneous agent model. However, the heterogeneity in the types of firms can make it such that these effects are greatly mitigated in the total network density.

\begin{figure}
    \begin{center}
    \caption{Effect of taxes on network density}%
    \subfloat[Total network density]{{\includegraphics[width=0.49\linewidth]{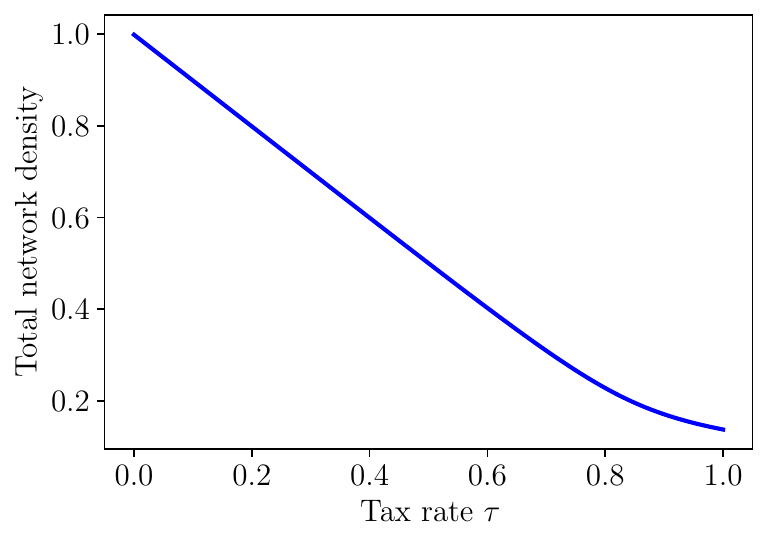} }}%
    \subfloat[Density kernel]{{\includegraphics[width=0.49\linewidth]{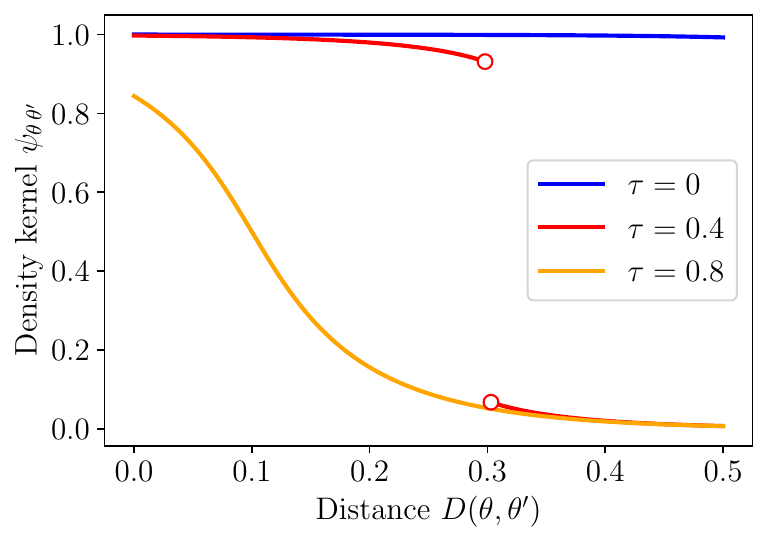} }}
    \label{fig:tax_phase_transition}
    \end{center}
    \footnotesize{\textit{Notes.} This figure shows the effects of changing the tax rate $\tau$ on network density for $L = 100$ types and model parameters $(\gamma, r) = (10,10)$. Panel (a) shows the behavior of the total network density as $\tau$ is varied. Panel (b) shows how the density kernel as a function of distance between types changes for different tax rates.}
\end{figure}

To build intuition on this result, consider the case where the limiting distribution ${\bf w}$ is the uniform distribution over $\Theta$. The total network density is $\rho \equiv \sum_{(\theta,\theta') \in \Theta^2} w_\theta w_{\theta'} \psi_{\theta \theta'}$. With this, we can study the differential effects of changing the tax rate on the total network density and on the density kernel, as shown in Figure \ref{fig:tax_phase_transition}. In this case, increasing the tax rate $\tau$ decreases the total density of the network with almost imperceptible jumps. In contrast, the density kernel for any two given types can exhibit a stark discontinuous transition. In this example, we see that understanding the global phase transition structure in the motif-only model gives us insight into the mechanisms of these ``local'' phase transitions.

\subsection{Complex trading structures}
In the previous model of trade, the firms could be thought of as thinking of a ``local'' trade-off, where the only factor that mattered to form a link was whether it would create a mutual link or not. We can think of more complex incentive structures, in which there is feedback from the structure of the rest of the network.

Consider a supply-chain model, where firms receive a revenue $r/N^{\ell-1}$ if they participate in an $\ell$-link chain, defined as the motif $c_\ell \equiv \{12,23,\ldots,(\ell-1)\ell\}$. Firms of type $\theta$ still need to pay a fixed cost of $\gamma D(\theta, \theta')$ to connect to firms of type $\theta'$. Using Theorem \ref{thm:mult_types_partition}, we can characterize the typical limiting behavior in this model.

\begin{lemma} \label{lem:trade_global_fixed_point}
    Suppose that the limiting distribution ${\bf w}$ is the uniform distribution over $\Theta$. Then the maximizer $\psi^*$ for the corresponding optimization problem is given by
    \begin{align}
        \psi^*_{\theta \theta'} = \left[ 1 + \exp(\gamma D(\theta,\theta') - \ell r (\rho^*)^{\ell - 1} ) \right]^{-1},
    \end{align}
    where $\rho^*$ is a fixed point of the function
    \begin{align}
        R(\rho;\gamma,r,\ell) \equiv \frac{1}{L^2} \sum_{\theta,\theta' \in \Theta^2} \left[ 1 + \exp(\gamma D(\theta,\theta') - \ell r \rho^{\ell - 1}) \right]^{-1}.
    \end{align}
\end{lemma}

Lemma \ref{lem:trade_global_fixed_point} shows that there is a very similar mechanism at play in this setting as in the model with only motif values. The typical density fo the network $\rho^*$ is determined by a fixed point problem that takes into account the contributions from both the utility and entropy terms. The kernel $\psi^*_{\theta \theta'}$ is then determined by the typical density of the network and the individual incentives that firms of type $\theta$ have to form trade routes with firms of type $\theta'$. This gives a clear characterization of the interplay between global network density and the composition of the neighborhoods of individual firms.

\begin{figure}
    \begin{center}
    \caption{Total density in the supply chain model}
    \includegraphics[width = 0.6 \linewidth]{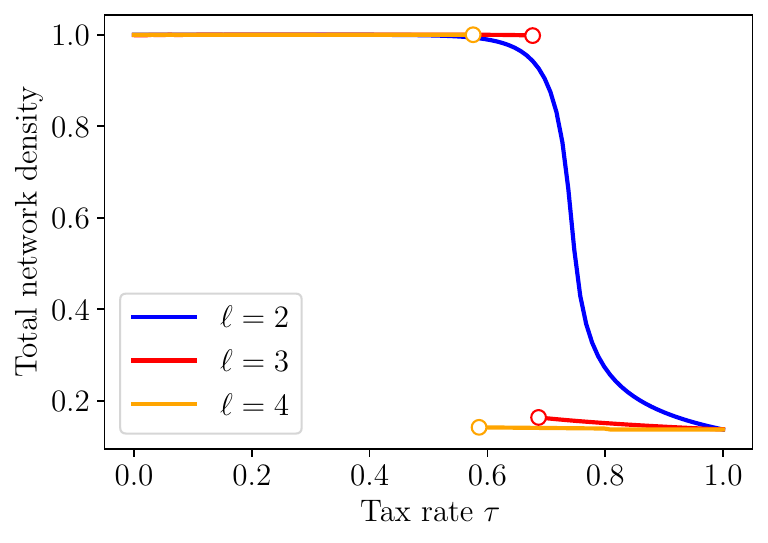}
    \label{fig:supply_chain_density}
    \end{center}
    \footnotesize{\textit{Notes.} This figure shows the effect of changing the tax rate on the typical density for $L=100$ types and model parameters $(\gamma, r) = (10,10)$ for various lenghts of the supply chain.}
\end{figure}

The effect of taxes on total network density is shown in Figure \ref{fig:supply_chain_density} for different lengths of the supply chain. An interesting phenomenon occurs as the length of the supply chain motif is increased: the decrease in network density can go from being continuous to exhibiting a discontinuous phase transition. This is a similar mechanism to the appearance of a phase transition in Figure \ref{fig:star_phase_transition}, where sufficient complexity in the motif is necessary to generate discontinuities. 

This simple example illustrates that, depending on the structure of incentives, phase transitions can take on completely different forms. Without the feedback mechanisms present in the mutual-links motif $m_2$, there is no ``local'' phase transition in this model, but it still have global phase transitions depending on the complexity of the chain.
\section{Discussion} \label{sec:conclusion}
In this paper I show that under a condition called \textit{conservativeness}, the stationary distribution of a stochastic best-response network formation game converges to a Gibbs measure with an potential that is constructed from the deterministic components of agents' utilities. Using the theory of potential games, I identify the local maxima of the stationary distribution with Nash equilibria of the deterministic game.

The results on aggregation functions in this paper apply to general utility functions. However, the existence of potentials associated to this network formation game is strongly related to the reversibility of the process, and the interpretation of the stationary distribution as the generator of long-run averages relies on the process having sufficient time to explore a significant fraction of network space, which can be difficult to justify for social interactions with many agents. Exploring results beyond the reversibility of the process and establishing the appropriate timescales for their application is a topic left for future research.

For games that allow for the construction of a potential, I identify the normalization constant in the corresponding Gibbs measure with the partition function from statistical physics. This implies that the normalization constant encodes information about the long-run average of the derivatives of the aggregation function with respect to model parameters, which correspond to sufficient statistics of the network in the case of Exponential Random Graph Models (ERGMs). For the case of motif utilities, I establish a trade-off between incentives and the exponentially increasing size of network space in establishing the asymptotic behavior of the partition function and the statistical properties of the model. 

To illustrate the forces in this model, I apply this methodology to a simple model of establishment of trade routes. I show that the model exhibits phase transitions when changing the tax rate on trade, meaning that a continuous change in the tax rate can lead to a discontinuous change in network density. In addition, the complexity of the trading structures that are formed plays a role in the emergence of these phase transitions. While these results rely on having an infinite amount of players, it is expected that for finite but large $N$ there would be similar sharp changes in the properties of the model. 

\setstretch{1}
\bibliographystyle{aer}
\bibliography{references}
\setstretch{1.5}

\newpage

\appendix

\setcounter{page}{1}
\renewcommand{\thesubsection}{\thesection.\arabic{subsection}}
\renewcommand{\thefootnote}{\hbAppendixPrefix\arabic{footnote}}
\setcounter{footnote}{0}
\renewcommand{\thefigure}{\hbAppendixPrefix\Roman{figure}}
\setcounter{figure}{0}
\renewcommand{\thetable}{\hbAppendixPrefix\Roman{table}}
\setcounter{table}{0}
\renewcommand{\theequation}{\hbAppendixPrefix\arabic{equation}}
\setcounter{equation}{0}

\begin{center} \huge \textbf{Appendix} \end{center}
\section{Proofs} \label{sec:app_proofs}
\subsection{Proof of Proposition \ref{prop:conv}}
To prove this result, it is useful to establish the following technical lemma:
\begin{lemma} \label{lem:conservative_cost}
    Consider a function $\phi:{\cal D} \times {\cal G} \to \mathbb{R}$ with the property that $\phi_{ij}(g) = - \phi_{ij}(\sigma_{ij}(g))$. The function satisfies
    \begin{align} \label{eq:conservative_cost}
        \phi_{ij}(g) + \phi_{i'j'}(\sigma_{ij}(g)) = \phi_{i'j'}(g) + \phi_{ij}(\sigma_{i'j'}(g))
    \end{align}
    for all $ij, i'j' \in {\cal D}$ and $g \in {\cal G}$ if and only if there exists a function $\Psi: {\cal G} \to \mathbb{R}$ such that
    \begin{align} \label{eq:cost_potential}
        \phi_{ij}(g) = \Psi(\sigma_{ij}(g)) - \Psi(g)
    \end{align}
    for all $ij \in {\cal D}$ and $g \in {\cal G}$. Furthermore, for a given $\phi$, this function $\Psi$ is unique up to an additive constant.
\end{lemma}

\begin{proof}
    The ``if'' direction can be immediately obtained by direct computation of Equation \eqref{eq:conservative_cost} using the expression for $\phi$ in terms of $\Psi$.

    The ``only if'' direction can be proven by construction. Fix a network $g \in {\cal G}$ and let $n = |g|$ be its size. Let $\gamma = (\gamma_1, \ldots, \gamma_n)$ be an ordering of the links in $g$, meaning that $\gamma_\ell \in g$ for $\ell = 1, \ldots, n$ and $\gamma_\ell \ne \gamma_{\ell'}$ if $\ell \ne \ell'$. From this ordering, we can construct a sequence of networks recursively by letting $g^\gamma_0 \equiv \varnothing$ and setting $g^\gamma_{\ell} \equiv \sigma_{\gamma_\ell}(g^\gamma_{\ell-1})$ for $\ell > 0$. Let us define the following quantity:
    \begin{align}
        \Psi^\gamma(g) = \sum_{\ell = 1}^n \phi_{\gamma_\ell}(g^\gamma_{\ell-1}).
    \end{align}
    I will now prove that this value does not depend on the specific ordering of the links chosen. 
    
    First, let us consider an ordering $\gamma^0$ that differs from $\gamma$ only at positions $\ell_0$ and $\ell_0+1$, where $1 \le \ell_0 < n$. Since both $\gamma$ and $\gamma^0$ are orderings, this means that $\gamma^0$ is $\gamma$ with the entries $\ell_0$ and $\ell_0+1$ swapped. From this, we must have $g^{\gamma_0}_{\ell_0-1} = g^{\gamma}_{\ell_0-1}$, and $g^{\gamma_0}_{\ell_0+2} = g^{\gamma}_{\ell_0+2}$ if $\ell_0 < n-1$. Therefore, computing $\Psi^\gamma(g) - \Psi^{\gamma^0}(g)$ yields
    \begin{align}
        \Psi^\gamma(g) - \Psi^{\gamma^0}(g) = \phi_{\gamma_{\ell_0}}(g^\gamma_{\ell_0-1}) + \phi_{\gamma_{\ell_0+1}}(g^\gamma_{\ell_0}) - \phi_{\gamma^0_{\ell_0}}(g^{\gamma^0}_{\ell_0-1}) - \phi_{\gamma^0_{\ell_0+1}}(g^{\gamma^0}_{\ell_0}).
    \end{align}
    Note that we can write $g^{\gamma^0}_{\ell_0} = \sigma_{\gamma_{\ell_0+1}}(g^{\gamma}_{\ell_0 - 1})$. Therefore, this difference can be written as
    \begin{align}
        \Psi^\gamma(g) - \Psi^{\gamma^0}(g) = \phi_{\gamma_{\ell_0}}(g^\gamma_{\ell_0-1}) + \phi_{\gamma_{\ell_0+1}}(\sigma_{\gamma_{\ell_0}}(g^\gamma_{\ell_0-1})) - \phi_{\gamma_{\ell_0+1}}(g^{\gamma}_{\ell_0-1}) - \phi_{\gamma_{\ell_0}}(\sigma_{\gamma_{\ell_0+1}}(g^{\gamma}_{\ell_0 - 1})).
    \end{align}
    From the condition in Equation \eqref{eq:conservative_cost}, this must vanish, such that $\Psi^\gamma(g) = \Psi^{\gamma^0}(g)$. Any ordering of the links in $g$ can be obtained starting from $\gamma$ through a sequence of ``swaps'' like the one we just analyzed. Therefore, the quantity $\Psi^\gamma(g)$ will be the same regardless of the ordering $\gamma$ chosen. Thus, we can define a quantity that depends solely of the network $\Psi(g) \equiv \Psi^\gamma(g)$ for any ordering $\gamma$.

    Now we want to prove that our construction of $\Psi$ satisfies Equation \eqref{eq:cost_potential}. Fix a network $g$ and a link $ij \in {\cal D}$. Without loss of generality, suppose that $ij \ne g$\footnote{Since $\phi_{ij}(g) = - \phi_{ij}(\sigma_{ij}(g))$, it suffices to prove that Equation \eqref{eq:cost_potential} holds for $g$ to prove that it holds for $\sigma_{ij}(g)$, so we can choose to prove it for the smaller network.}. Let $\gamma^1$ be some ordering of the links in $g$ and $\gamma^2$ be the ordering of the links in $\sigma_{ij}(g)$ such that $\gamma^2_\ell = \gamma^1_\ell$ for $0 \le \ell \le |g|$ and $\gamma^2_{|g|+1} = ij$. We can write our $\Psi$ functions using these orderings, such that
    \begin{align}
        \Psi(\sigma_{ij}(g)) - \Psi(g) = \sum_{\ell = 1}^{|g|+1} \phi_{\gamma^2_\ell}(g^{\gamma^2}_{\ell-1}) - \sum_{\ell = 1}^{|g|} \phi_{\gamma^1_\ell}(g^{\gamma^1}_{\ell-1}) = \phi_{ij}(g),
    \end{align}
    as desired.

    Finally, to prove uniqueness up to an additive constant, suppose that there are two functions $\Psi_1$ and $\Psi_2$ satisfying Equation \eqref{eq:cost_potential} for all $g \in {\cal G}$ and $ij \in {\cal D}$. Let $K_0 \equiv \Psi_1(\varnothing) - \Psi_2(\varnothing)$, and suppose there exists a $g_0$ such that $\Psi_1(g_0) - \Psi_2(g_0) \ne K_0$. From Equation \eqref{eq:cost_potential}, we have that, for all $g \in {\cal G}$ and $ij \in {\cal D}$
    \begin{align}
        \Psi_1(g) - \Psi_2(g) &= [\Psi_1(g) + \phi_{ij}(g)] - [\Psi_2(g) + \phi_{ij}(g)] \nonumber \\
        &= \Psi_1(\sigma_{ij}(g)) - \Psi_2(\sigma_{ij}(g)).
    \end{align}
    Since this procedure can be repeated with any sequence of dyads to reach any network, we conclude that $\Psi_1(g) - \Psi_2(g) = K_0$ for all networks, which is a contradiction. 
\end{proof}

We can now prove Proposition \ref{prop:conv}. We can first prove that utilities are conservative if and only if the potential exists. To do this, define $\phi_{ij}(g) \equiv V_i(\sigma_{ij}(g)) - V_i(g)$. Conservativeness, then, is equivalent to Equation \eqref{eq:conservative_cost}, so from Lemma \ref{lem:conservative_cost}, we have the equivalence between the first two statements of the Proposition.

The equivalence with the third statement is proven in two parts:
\begin{itemize}
    \item \textit{Existence of the potential implies reversibility of the process:} By definition, the potential satisfies
    \begin{align}
        \Phi(\sigma_{ij}(g))-\Phi(g) = V_i(\sigma_{ij}(g)) - V_i(g) = \log \left( \frac{p_{ij}(g)}{p_{ij}(\sigma_{ij}(g))} \right).
    \end{align}
    The Markov process is reversible if there exists a distribution $\pi$ such that the detailed balance condition is satisfied:
    \begin{align}
        \lambda_{ij}(g) p_{ij}(g) \pi(g) = \lambda_{ij}(\sigma_{ij}(g)) p_{ij}(\sigma_{ij}(g)) \pi(\sigma_{ij}(g))
    \end{align}
    for all $g \in {\cal G}$ and $ij \in {\cal D}$. This is sufficient since the only non-zero transition probabilities correspond to pairs of networks that differ by only one link. Since $\lambda_{ij}(\sigma_{ij}(g)) = \lambda_{ij}(g)$, we have that
    \begin{align}
        \lambda_{ij}(g) p_{ij}(g) e^{\Phi(g)} &= \lambda_{ij}(\sigma_{ij}(g)) p_{ij}(\sigma_{ij}(g)) \left[ \frac{p_{ij}(g)}{p_{ij}(\sigma_{ij}(g))} \right] e^{\Phi(g)} \nonumber \\
        &= \lambda_{ij}(\sigma_{ij}(g)) p_{ij}(\sigma_{ij}(g)) e^{\Phi(\sigma_{ij}(g))-\Phi(g)} e^{\Phi(g)} \nonumber \\
        &= \lambda_{ij}(\sigma_{ij}(g)) p_{ij}(\sigma_{ij}(g)) e^{\Phi(\sigma_{ij}(g))}.
    \end{align}
    Define the following probability distribution over the space of networks:
    \begin{align} \label{eq:stationary_dist}
        \Tilde{\pi}(g) = \frac{e^{\Phi(g)}}{\sum_{g' \in {\cal G}} e^{\Phi(g')}}.
    \end{align}
    Dividing the previous expression by $\sum_{g' \in {\cal G}} e^{\Phi(g')}$, we have that
    \begin{align}
        \lambda_{ij}(g) p_{ij}(g) \tilde{\pi}(g) = \lambda_{ij}(\sigma_{ij}(g)) p_{ij}(\sigma_{ij}(g)) \tilde{\pi}(\sigma_{ij}(g)),
    \end{align}
    so the process is reversible. Additionally, satisfying detailed balance implies that $\tilde{\pi}$ is the unique stationary distribution of the Markov chain.

    \item \textit{Reversibility of the process implies existence of the potential:} Recall that under Assumption \ref{A:T1EV}, there exists a unique non-degenerate stationary distribution $\pi$ to which the system converges. If the process is reversible, it satisfies the detailed balance condition:
    \begin{align}
        \lambda_{ij}(g) p_{ij}(g) \pi(g) &= \lambda_{ij}(\sigma_{ij}(g)) p_{ij}(\sigma_{ij}(g)) \pi(\sigma_{ij}(g))
    \end{align}
    Since the stationary distribution has full support and $\lambda_{ij}(g) = \lambda_{ij}(\sigma_{ij}(g))$, we have that
    \begin{align}
        \log(\pi(\sigma_{ij}(g))) - \log(\pi(g)) &= \log \left( \frac{p_{ij}(g)}{p_{ij}(\sigma_{ij}(g))} \right).
    \end{align}
    for all $g \in {\cal G}$ and $ij \in {\cal D}$. Since the stationary probabilities are a function of the network only, we have that any function
    \begin{align}
        \Phi(g) = C + \log(\pi(g)),
    \end{align}
    with $C \in \mathbb{R}$, will satisfy
    \begin{align}
        \Phi(\sigma_{ij}(g)) - \Phi(g) = \log\left( \frac{p_{ij}(g)}{p_{ij}(\sigma_{ij}(g))} \right) = V_i(\sigma_{ij}(g)) - V_i(g).
    \end{align}
\end{itemize}
Using the proof of the first implication, we have that the potential characterizes the stationary distribution of the chain. Namely, the stationary distribution is given by Equation \eqref{eq:stationary_dist}.

\subsection{Proof of Proposition \ref{prop:value_sharing}}
I prove that the value-sharing representation can be constructing via induction on the number of links in the network.

We can always define $\hat{V}_i(g) = V_i(g) - V_i(g^{-i})$ with the same potential. Note that
\begin{align} \label{eq:utility_potential_representation}
    \Phi(g) - \Phi(g^{-i}) = \hat{V}_i(g) - \hat{V}_i(g^{-i}) = \hat{V}_i(g).
\end{align}

We proceed by induction on the size of the network.

What we want to show is that
\begin{align}
    \hat{V}_i(g) - \sum_{g_0 \subset g} \alpha(g_0) \mathbbm{1}\{i \in {\cal N}^{\textrm{src}}(g_0)\} = \hat{V}_j(g) - \sum_{g_0 \subset g} \alpha(g_0) \mathbbm{1}\{j \in {\cal N}^{\textrm{src}}(g_0)\} 
\end{align}
for all $i,j \in {\cal N}^{\textrm{src}}(g)$. If $i = j$, this is clearly true, so we focus on the case $i \ne j$. We can see that the sums will only differ in the subgraphs $g_0$ for which only one of $i$ and $j$ is present. Therefore, we can re-write this expression as
\begin{align}
    \hat{V}_i(g) - \sum_{g_0 \subseteq g^{-j}} \alpha(g_0) \mathbbm{1}\{i \in {\cal N}^{\textrm{src}}(g_0)\} = \hat{V}_j(g) - \sum_{g_0 \subseteq g^{-i}} \alpha(g_0) \mathbbm{1}\{j \in {\cal N}^{\textrm{src}}(g_0)\},
\end{align}
where we no longer need a strict subset relation since $g^{-i} \ne g$ and $g^{-j} \ne g$. By our induction assumption, these sums correspond to $\hat{V}_i(g^{-j})$ and $\hat{V}_j(g^{-i})$, respectively, such that our condition is
\begin{align}
    \hat{V}_i(g) - \hat{V}_i(g^{-j}) = \hat{V}_j(g) - \hat{V}_j(g^{-i}).
\end{align}
Letting $g^{-i-j} \equiv (g^{-i})^{-j} = (g^{-j})^{-i}$, we can write this condition in terms of the potential using Equation \eqref{eq:utility_potential_representation}:
\begin{align}
    \Phi(g) - \Phi(g^{-i}) - [\Phi(g^{-j}) - \Phi(g^{-i-j})] = \Phi(g) - \Phi(g^{-j}) - [\Phi(g^{-i}) - \Phi(g^{-i-j})],
\end{align}
which is clearly satisfied.

The uniqueness of the coefficients comes from the fact that they can be obtained from a M\"obius inversion of the potential. Specifically, if we define
\begin{align}
    \alpha(g) = \sum_{g_0 \subseteq g} (-1)^{|g \backslash g_0|} \Phi(g_0),
\end{align}
then these are the unique coefficients that satisfy
\begin{align}
    \Phi(g) = \sum_{g_0 \subseteq g} \alpha(g_0).
\end{align}

\subsection{Proof of Proposition \ref{prop:MPE}}
I will prove existence of a MPE using the Brouwer fixed point theorem. Consider the vector in ${\bf x} \equiv ((x_{i}(g))_{i \in {\cal N}})_{g \in {\cal G}} \in \mathbb{R}^{N \times |{\cal G}|}$, which represents some set of values that agents assign to networks. For a given $\rho > 0$, define the function $T: \mathbb{R}^{N \times |{\cal G}|} \to \mathbb{R}^{N \times |{\cal G}|}$ by
\begin{align}
    [T({\bf x})]_i(g) &= \sum_{g' \in {\cal G}} v_i(g') \Tilde{\mathbb{P}}(g',g \,|\, \rho, {\bf x}),
\end{align}
where
\begin{align}
    \Tilde{\mathbb{P}}(g',g \,|\, \rho, {\bf x}) = \rho \int_0^\infty e^{-\rho t} \mathbb{P}\{G_t = g' \, | \, G_0 = g, {\bf x} \} \, dt,
\end{align}
and $\mathbb{P}\{G_t = g' \, | \, G_0 = g, {\bf x} \}$ is the probability that the network at time $t$ is $g'$ given that the network at time $0$ is $g$, assuming the dynamics are generated by the value functions encoded in ${\bf x}$. Note that these time-weighted probabilities satisfy
\begin{align}
    \sum_{g' \in {\cal G}} \Tilde{\mathbb{P}}(g',g \,|\, \rho, {\bf x}) &= \rho \int_0^\infty e^{-\rho t} \left[ \sum_{g' \in {\cal G}} \mathbb{P}\{G_t = g' \, | \, G_0 = g, {\bf x} \} \right] \, dt \nonumber \\
    &= \rho \int_0^\infty e^{-\rho t} \, dt \nonumber \\
    &= 1.
\end{align}
As such, we have that $[T({\bf x})]_i(g)$ is a convex combination of the $v_i(g)$ across all networks, for any value of ${\bf x}$. Define
\begin{align}
    \underline{v}_i \equiv \min_{g \in {\cal G}} v_i(g), \quad \overline{v}_i \equiv \max_{g \in {\cal G}} v_i(g).
\end{align}
We, therefore, have that the image of $T$ is ${\cal V} \equiv (\prod_{i \in {\cal N}}[\underline{v}_i, \overline{v}_i])^{|{\cal G}|}$. Restricting $T$ to the domain ${\cal V} \subset \mathbb{R}^{N \times |{\cal G}|}$, we conclude that $T$ maps ${\cal V}$ to itself. 

I now prove that $T$ is continuous in ${\bf x}$. We have that the switching probabilities associated to a given ${\bf x}$ are given by
\begin{align}
    p_{ij}(g \, | \, {\bf x}) &= [1 + \exp(x_i(g) - x_i(\sigma_{ij}(g)))]^{-1}.
\end{align}
Clearly, the switching probabilities are differentiable in ${\bf x}$. The probability distribution on ${\cal G}$ satisfies the Kolmogorov forward equation
\begin{align}
    \dot{\mathbb{P}}\{G_t = g' \, | \, G_0 = g, {\bf x} \} = \sum_{ij \in {\cal D}} \lambda_{ij}(g') \Bigg[ & p_{ij}(\sigma_{ij}(g') | \, {\bf x}) \mathbb{P}\{G_t = \sigma_{ij}(g') \, | \, G_0 = g, {\bf x} \} \nonumber \\
    & - p_{ij}(g' \, | \, {\bf x}) \mathbb{P}\{G_t = g' \, | \, G_0 = g, {\bf x} \} \Bigg],
\end{align}
with initial conditions
\begin{align}
    \mathbb{P}\{G_0 = g' \, | \, G_0 = g, {\bf x} \} &= \mathbbm{1}\{g'=g\}.
\end{align}
Define the function $S:{\cal G} \times [0,1]^{|{\cal G}|} \times {\cal V} \to \mathbb{R}$ by
\begin{align}
    S(g,{\bf y} \, | \, {\bf x}) \equiv \sum_{ij \in {\cal D}} \lambda_{ij}(g) [ p_{ij}(\sigma_{ij}(g) | \, {\bf x}) y(\sigma_{ij}(g))  - p_{ij}(g \, | \, {\bf x}) y(g) ].
\end{align}
Since the switching probabilities are differentiable in ${\bf x}$, we have that $S$ is differentiable in ${\bf x}$. Additionally, since the switching probabilities are bounded, we also have that $S$ is differentiable in ${\bf y}$. We, therefore, have that the function is Lipschitz. Using Theorem 8.3 in \citet{Amann1990OrdinaryEquations}, we conclude that the solution of the differential equation $\mathbb{P}\{G_t = g \, | \, G_0 = g, {\bf x} \}$ is continuous in ${\bf x}$. Using continuity of the solution, we have that the integral $\tilde{\mathbb{P}}\{g, g' \, | \, \rho, {\bf x} \}$ is continuous in ${\bf x}$, according to Theorem 5.6 in \citet{Elstrodt1996Ma-Integrationstheorie}. This implies that $T$ is continuous in ${\bf x}$.

Since $T$ is continuous and maps a convex compact subset of $\mathbb{R}^{N \times |{\cal G}|}$ to itself, by Brouwer's fixed-point theorem, $T$ must have a fixed point. That is, there exists an ${\bf x}^* \in {\cal V}$ such that
\begin{align}
    {\bf x}^*_i(g) = \sum_{g' \in {\cal G}} v_i(g') \Tilde{\mathbb{P}}(g',g \,|\, \rho, {\bf x}^*).
\end{align}
As such, a MPE exists for any $\rho > 0$.

\subsubsection{$\rho_n \to \infty$ limit}
Note that the value functions in a MPE corresponding to a discount factor $\rho$ must satisfy the Hamilton-Jacobi-Bellman equation:
\begin{align} \label{eq:hjb}
    \rho V_i(g) = \rho v_i(g) + \sum_{i'j' \in {\cal D}} \lambda_{i'j'}(g) p_{i'j'}(g) [V_i(\sigma_{i'j'}(g)) - V_i(g)]
\end{align}
for all $g \in {\cal G}$ and $i \in {\cal N}$. This implies that, for any $n$ in our sequence of equilibria, the following holds:
\begin{align}
    |V^n_i(g) - v_i(g)| &= \frac{1}{\rho_n} \left| \sum_{i'j' \in {\cal D}} \lambda_{i'j'}(g) p^n_{i'j'}(g) [V^n_i(\sigma_{i'j'}(g)) - V^n_i(g)] \right| \nonumber \\
    &\le \frac{1}{\rho_n} \sum_{i'j' \in {\cal D}} \lambda_{i'j'}(g) p^n_{i'j'}(g) (|V^n_i(\sigma_{i'j'}(g))| + |V^n_i(g)|).
\end{align}
Note that the meeting rates do not depend on the equilibrium and the switching probabilities must be in $(0,1)$. As was shown above, the fixed points of the function $T$ lie in a bounded set that is independent of $n$, so the sum on the right is bounded by some constant $K_1$, which is independent of $n$:
\begin{align}
    |V^n_i(g) - v_i(g)| \le \frac{K_1}{\rho_n}.
\end{align}
Therefore, we see that for any sequence of equilibria for which $\rho_n \to \infty$, we must have
\begin{align}
    \lim_{n \to \infty} |V^n_i(g) - v_i(g)| = 0.
\end{align}

\subsubsection{$\rho_n \to 0$ limit}
Since for every value function the transition probabilities are bounded away from 0, we have a finite mixing time. Define
\begin{align}
    T(\epsilon) \equiv \sup_{{\bf V}}\{\textrm{mixing time for probabilities given by } {\bf V} \textrm{ with threshold } \epsilon\}.
\end{align}
We can write the value functions as a contribution before $T(\epsilon)$ and a contribution after:
\begin{align}
    V_i^n(g) =& \rho_n \int_0^{T(\epsilon)} \sum_{g' \in {\cal G}} v_i(g') e^{-\rho_n t} \mathbb{P}\{G_t = g' \, | \, G_0 = g, {\bf V}^n \} \, dt \nonumber \\
    &+ \rho_n \int_{T(\epsilon)}^\infty \sum_{g' \in {\cal G}} v_i(g') e^{-\rho_n t} \mathbb{P}\{G_t = g' \, | \, G_0 = g, {\bf V}^n \} \, dt.
\end{align}
Define the weighted average of the flow values under the stationary distribution induced by ${\bf V}^n$ as
\begin{align}
    \tilde{V}_i^n \equiv \sum_{g' \in {\cal G}} \pi(g' \, | \, {\bf V}^n) v_i(g') = \rho_n \int_0^\infty \sum_{g' \in {\cal G}} v_i(g') e^{-\rho_n t} \pi(g' \, | \, {\bf V}^n).
\end{align}
Subtracting this value from the above expression, we obtain
\begin{align}
    V_i^n(g) - \tilde{V}_i^n =& \rho_n \int_0^{T(\epsilon)} \sum_{g' \in {\cal G}} v_i(g') e^{-\rho_n t} [\mathbb{P}\{G_t = g' \, | \, G_0 = g, {\bf V}^n \} - \pi(g' \, | \, {\bf V}^n)] \, dt \nonumber \\
    &+ \rho_n \int_{T(\epsilon)}^\infty \sum_{g' \in {\cal G}} v_i(g') e^{-\rho_n t} [\mathbb{P}\{G_t = g' \, | \, G_0 = g, {\bf V}^n \} - \pi(g' \, | \, {\bf V}^n)] \, dt.
\end{align}
By definition of $T(\epsilon)$, the mixing time of the dynamics under ${\bf V}^n$ is bounded by $T(\epsilon)$. Additionally, by the definition of mixing time, for all $t \ge T(\epsilon)$ and all $g' \in {\cal G}$, we have that
\begin{align}
    |\mathbb{P}\{G_t = g' \, | \, G_0 = g, {\bf V}^n \} - \pi(g' \, | \, {\bf V}^n)| \le \epsilon.
\end{align}
This implies that
\begin{align}
    \int_{T(\epsilon)}^\infty e^{-\rho_n t} |\mathbb{P}\{G_t = g' \, | \, G_0 = g, {\bf V}^n \} - \pi(g' \, | \, {\bf V}^n)| \, dt \le \epsilon \int_{T(\epsilon)}^\infty e^{-\rho_n t} \, dt \le \epsilon \int_{0}^\infty e^{-\rho_n t} \, dt = \frac{\epsilon}{\rho_n}.
\end{align}
This yields the following bound:
\begin{align}
    |V_i^n(g) - \tilde{V}_i^n| \le \sum_{g' \in {\cal G}} |v_i(g')| \left[ \epsilon +  \rho_n \int_0^{T(\epsilon)} e^{-\rho_n t} |\mathbb{P}\{G_t = g' \, | \, G_0 = g, {\bf V}^n \} - \pi(g' \, | \, {\bf V}^n)| \, dt \right].
\end{align}
Now, note that $|\mathbb{P}\{G_t = g' \, | \, G_0 = g, {\bf V}^n \} - \pi(g' \, | \, {\bf V}^n)| \le 1$, since both quantities are probabilities, such that
\begin{align}
    \rho_n \int_0^{T(\epsilon)} e^{-\rho_n t} |\mathbb{P}\{G_t = g' \, | \, G_0 = g, {\bf V}^n \} - \pi(g' \, | \, {\bf V}^n)| \, dt \le 1 - e^{- \rho_n T(\epsilon)}.
\end{align}
Therefore, our bound now is
\begin{align}
    |V_i^n(g) - \tilde{V}_i^n| \le \sum_{g' \in {\cal G}} |v_i(g')| \left[ \epsilon + 1 - e^{- \rho_n T(\epsilon)} \right].
\end{align}
For sufficiently large $n$, the term $1 - e^{- \rho_n T(\epsilon)}$ can be made arbitrarily small, since $\rho_n \to 0$. Namely, it can be made smaller than $\epsilon$. Therefore, for large $n$, we have
\begin{align}
    |V_i^n(g) - \tilde{V}_i^n| \le 2 \epsilon \sum_{g' \in {\cal G}} |v_i(g')|.
\end{align}
Since this bound holds for all $g$, it must be true that the distance between the values for any two networks is also bounded:
\begin{align}
    |V_i^n(g) - V_i^n(g')| \le 4 \epsilon \sum_{g'' \in {\cal G}} |v_i(g'')| \quad \forall g, g' \in {\cal G}.
\end{align}
Therefore, since $\epsilon$ can be made arbitrarily small, we must have that
\begin{align}
    \lim_{n \to \infty} |V_i^n(g) - \tilde{V}_i^n| &= 0 \quad \forall g \in {\cal G}, \nonumber \\
    \lim_{n \to \infty} |V_i^n(g) - V_i^n(g')| &= 0 \quad \forall g, g' \in {\cal G}.
\end{align}

To finish the proof, note that the switching probabilities are continuous functions of differences in the values of different networks. From the result above, this means that
\begin{align}
    \lim_{n \to \infty} p_{ij}(g \, | \, {\bf V}^n) = \frac{1}{2} \quad \forall ij \in {\cal D}, g \in {\cal G}.
\end{align}
Additionally, note that the Markov chain is irreducible for any ${\bf V}$. This means that the stationary distribution is a continuous function of the transition rates (and, hence, the switching probabilities), since this distribution is determined by a linear system of equations whose coefficients are the rates. If all the switching probabilities were $1/2$, the stationary distribution would be uniform, so by continuity we must have
\begin{align}
    \lim_{n \to \infty} \pi(g \, | \, {\bf V}^n) = \frac{1}{|{\cal G}|} \quad \forall g \in {\cal G}.
\end{align}
Therefore, the limit of the auxiliary values $\tilde{V}_i^n$ exists, and is given by
\begin{align}
    \lim_{n \to \infty} \tilde{V}^n_i = \frac{1}{|{\cal G}|} \sum_{g' \in {\cal G}} v_i(g').
\end{align}
Since this limit exists, we must have that, for all $g$,
\begin{align}
    \lim_{n \to \infty} V^n_i(g) = \lim_{n \to \infty} \tilde{V}^n_i = \frac{1}{|{\cal G}|} \sum_{g' \in {\cal G}} v_i(g'),
\end{align}
as desired.

\subsection{Proof of Proposition \ref{prop:motif_partition}}
For definitions and important results, see Appendix \ref{sec:app_graph_limits}.

First, we want to find the set of graphons that solve the optimization problem in Theorem \ref{thm:partition_func_convergence}. Note that in our case, the function ${\cal T}$ corresponds to our scaled potential. These technically correspond to subgraph densities but, as pointed out in Appendix \ref{sec:app_graph_limits}, homomorphism densities and subgraph densities are equivalent for large networks. Specifically, in the proofs of Theorem \ref{thm:partition_func_convergence} and \ref{thm:graphon_convergence}, a subleading correction will be removed by the limsup and liminf. Therefore, we have that for a given graphon $h \in {\cal W}$,
\begin{align}
    {\cal T}(h) = \sum_{m \in {\cal M}} a_m \tilde{b}(m,h),
\end{align}
where $\tilde{b}(m,h)$ is the homomorphism density of motif $m$ into graphon $h$. To solve this problem, I will first prove that this optimization problem is solved by a constant graphon. 

Fix some graphon $h \in {\cal W}$ and a motif $m \in {\cal M}$. For our case of interest, H\"older's inequality states that for two functions $f,g:[0,1]^n \to [0,1]$ and for $p,q > 0$ such that $1/p +  1/q = 1$, we have that
\begin{align}
    \int_{[0,1]^n} f({\bf x}) g({\bf x}) \, dx_1 \ldots dx_n \le \left( \int_{[0,1]^n} f({\bf x})^p \, dx_1 \ldots dx_n \right)^{1/p} \left( \int_{[0,1]^n} g({\bf x})^q \, dx_1 \ldots dx_n \right)^{1/q},
\end{align}
with equality if and only if there exist $c_1, c_2 \in \mathbb{R}$, not both zero, such that $c_1 f({\bf x})^p = c_2 g({\bf x})^q$ almost everywhere. Applying this repeatedly with appropriate choices of $p$ and $q$, we have that for a collection of $k$ functions $f_i:[0,1]^n \to [0,1]$, the following holds:
\begin{align}
    \int_{[0,1]^n} \left( \prod_{i=1}^k f_i({\bf x}) \right) \, dx_1 \ldots dx_n \le \prod_{i=1}^k \left( \int_{[0,1]^n} f_i({\bf x})^k \, dx_1 \ldots dx_n \right)^{1/k}.
\end{align}
Applying this result to the homomorphism density of $m$ in $h$, we obtain
\begin{align}
    \tilde{b}(m,h) &= \int_{[0,1]^{n_m}} \left( \prod_{ij \in m} h(x_i, x_j) \right) dx_1 \ldots dx_{n_m} \nonumber \\
    &\le \prod_{ij \in m} \left( \int_{[0,1]^{n_m}} h(x_i, x_j)^{e_m} \, dx_1 \ldots dx_{n_m} \right)^{1/e_m} \nonumber \\
    &= \int_{[0,1]^2} h(x, y)^{e_m} \, dx \, dy,
\end{align}
where $e_m = |m|$ is the number of edges in the motif $m$. Additionally, note that the above is always an equality if $e_m = 1$.

Recall that we assumed $a_m > 0$ for all $m$ with $e_m > 1$. We can apply the previous result to the motifs with $e_m > 1$ to bound our function ${\cal T}$ for an arbitrary graphon $h$:
\begin{align}
    {\cal T}(h) \le \sum_{m \in {\cal M}} a_m \int_{[0,1]^2} h(x, y)^{e_m} \, dx \, dy.
\end{align}
Define the function $R:[0,1] \to \mathbb{R}$ as
\begin{align}
    R(s) \equiv \sum_{m \in {\cal M}} a_m s^{e_m} + H(s).
\end{align}
From the bound above, we see that
\begin{align}
    {\cal T}(h) + {\cal H}(h) \le \int_{[0,1]^2} R(h(x,y)) \, dx \, dy.
\end{align}
Let $R^* \equiv \sup_{s} R(s)$. Using the extreme value theorem, we know this supremum is actually a maximum, and is achieved at some $s^*$. Therefore, we have that
\begin{align}
    {\cal T}(h) + {\cal H}(h) \le \int_{[0,1]^2} R(s^*) \, dx \, dy.
\end{align}
Now, the graphon $h^*$ given by $h^*(x,y) = s^*$ saturates the H\"older bound and, therefore, satisfies
\begin{align}
    {\cal T}(h^*) + {\cal H}(h^*) = \int_{[0,1]^2} R(s^*) \, dx \, dy,
\end{align}
so it solves the optimization problem in Theorem \ref{thm:partition_func_convergence}.

Uniqueness of the solution follows the same argument as Theorem 4.1 in \citet{Chatterjee2013}. Therefore, convergence to the Erd\"os-R\'enyi model immediately follows from Theorem \ref{thm:graphon_convergence}.

\subsection{Proof of Theorem \ref{thm:mult_types_partition}}
To prove Theorem \ref{thm:mult_types_partition}, we want to find the colored graphon that solves the optimization problem in Theorem \ref{thm:colored_graphon_partition} in Appendix \ref{sec:app_graph_limits}. Intuitively, we want to prove that this problem is solved by a graphon that is piece-wise constant. 

Our potential is now composed of two terms ${\cal T}_1$ and ${\cal T}_2$, corresponding to the motif and neighbor utilities, respectively. These can be written as
\begin{align}
    {\cal T}_1(h,c) &= \sum_{m \in {\cal M}} a_m \tilde{b}(m,h), \nonumber \\
    {\cal T}_2(h,c) &= \sum_{\theta \in \Theta} \int_{[0,1]} \mathbbm{1}\{c(x)=\theta\} u_\theta \left( \left[ \int_{[0,1]} h(x,y) \mathbbm{1}\{c(y)=\theta'\} \, dy \right]_{\theta' \in \Theta} \right) \, dx.
\end{align}
To simplify notation, it is convenient to define ``subgraphons'' for every pair of types. Without loss of generality, we can consider the coloring to be ordered\footnote{this is because ${\cal T}$ is the same for all measurable re-labelings of the nodes.}, such that for some ordering of types $\theta^1, \ldots, \theta^L$, the coloring is
\begin{align}
    c(x) = \sum_{\theta^i \in \Theta} \theta^i \mathbbm{1}\left\{ x \ge \sum_{j < i} w_{\theta^j}, x < \sum_{j \le i} w_{\theta^j} \right\}.
\end{align}
The subgraphon associated to types $\theta^i$ and $\theta^j$ is, then, $h_{\theta^i \theta^j}: [0,1]^2 \to [0,1]$ defined by
\begin{align}
    h_{\theta^i \theta^j}(x,y) \equiv h\left( \sum_{k < i} w_{\theta^k} + w_{\theta^i} x, \sum_{k < j} w_{\theta^k} + w_{\theta^j} y \right).
\end{align}
This will allow us to consider the variational problem on the whole graphon as a variational problem on the subgraphons $h_{\theta \theta'}$.

To begin, we fix a colored graphon $(h,c)$ and find a bound on ${\cal T}_1(h)$ in a similar manner to before. For a given motif $m$, we can break up the integration domain into ``boxes'' where all the types are fixed. Specifically, we can write
\begin{align}
    \tilde{b}(m,h) &= \sum_{\boldsymbol{\theta} \in \Theta^{n_m}} \int_{[0,1]^{n_m}} \mathbbm{1}\{c(x_k) = \theta_k \, \forall k \} \left( \prod_{ij \in m} h(x_i, x_j) \right) \, dx_1 \ldots dx_{n_m} \nonumber \\
    &= \sum_{\boldsymbol{\theta} \in \Theta^{n_m}} \left( \prod_{i \in {\cal N}_m} w_{\theta_i} \right) \int_{[0,1]^{n_m}} \left( \prod_{ij \in m} h_{\theta_i \theta_j}(x_i, x_j) \right) \, dx_1 \ldots dx_{n_m}.
\end{align}
We will now use the following statement of H\"older's inequality for a product:
\begin{align}
    \int_{[0,1]^{n_m}} \left( \prod_{ij \in m} f_{ij}(x_i, x_j) \right) dx_1 \ldots dx_{n_m} \le \prod_{ij \in m} \lVert f_{ij} \rVert_{e_m}.
\end{align}
Applying H\"older's inequality to the homomorphism density, we obtain
\begin{align}
    \tilde{b}(m,h) &\le \sum_{\boldsymbol{\theta} \in \Theta^{n_m}} \left( \prod_{i \in {\cal N}_m} w_{\theta_i} \right) \prod_{ij \in m} \lVert h_{\theta_i \theta_j} \rVert_{e_m}.
\end{align}
We can use this to bound ${\cal T}_1$. Note that we can assume without loss of generality that $e_m > 1$ for all motifs $m$, since any motifs with $e_m = 1$ can be absorbed into the linear part of the neighborhood utility. Then, we have that $a_m > 0$ for all $m$, so we obtain the bound
\begin{align}
    {\cal T}_1(h,c) \le \sum_{m \in {\cal M}} a_m \sum_{\boldsymbol{\theta} \in \Theta^{n_m}} \left( \prod_{i \in {\cal N}_m} w_{\theta_i} \right) \prod_{ij \in m} \lVert h_{\theta_i \theta_j} \rVert_{e_m}.
\end{align}
Note that this bound has different norms based on the size of the motif. To obtain a bound that uses a single norm for all graphons, we can use the fact that for a space with $\lVert 1 \rVert_1 = 1$, for $1 \le p \le q < \infty$, the corresponding norms satisfy $\lVert h \rVert_p \le \lVert h \rVert_q$, with equality iff $h$ is constant almost everywhere. Define $e^* \equiv \max_{m \in {\cal M}} e_m$. Then we have the following bound:
\begin{align}
    {\cal T}_1(h,c) \le \sum_{m \in {\cal M}} a_m \sum_{\boldsymbol{\theta} \in \Theta^{n_m}} \left( \prod_{i \in {\cal N}_m} w_{\theta_i} \right) \prod_{ij \in m} \lVert h_{\theta_i \theta_j} \rVert_{e^*}.
\end{align}

To find a bound on ${\cal T}_2$, I will use the following technical Lemma.

\begin{lemma} \label{lem:entropy_bound}
    For any $n \ge 1$, $h \in {\cal W}$ and $c \in \mathbb{R}$,
    \begin{align}
        \int_{[0,1]^2} [H(h(x,y)) + c h(x,y)] \, dx \, dy \le H(\lVert h \rVert_{n}) + c \lVert h \rVert_{n}.
    \end{align}
\end{lemma}
\begin{proof}
    Note that the function $G(u) \equiv H(u) + c u$ has a unique maximum, which I call $\rho^*(c)$. I consider the cases $\lVert h \rVert_{n} \le \rho^*(c)$ and $\lVert h \rVert_{n} > \rho^*(c)$ separately.

    First, consider the case $\lVert h \rVert_{n} \le \rho^*(c)$. Since $G$ is a concave function, we can use Jensen's inequality to obtain
    \begin{align}
        \int_{[0,1]^2} G(h(x,y)) \, dx \, dy \le G(\lVert h \rVert_1).
    \end{align}
    Since $G(u)$ is non-decreasing for $u \le \rho^*(c)$ and $\lVert h \rVert_1 \le \lVert h \rVert_n$, we obtain the bound for this case.

    For the case $\lVert h \rVert_n > \rho^*(c)$, I show that the optimization problem
    \begin{align}
        \max_{h' \in {\cal W}} \int [H(h'(x,y)) + c h'(x,y)] \, dx \, dy \quad \textrm{s.t.} \quad \lVert h' \rVert_{n} \ge \rho,
    \end{align}
    for $\rho \equiv \lVert h \rVert_n > \rho^*(c)$, is solved by the constant graphon $h'(x,y) = \rho$. Note that $G$ satisfies
    \begin{align}
        \lim_{u \to 0} G'(u) = +\infty, \quad \lim_{u \to 1} G'(u) = -\infty.
    \end{align}
    This means that the constraints $h'(x,y) \in [0,1]$ will not be binding. Using Theorem 9.4.1 in \cite{luenberger1997optimization}, we have that the solution $h_n$ must be a stationary point of the Langrangian
    \begin{align}
        {\cal L}(h') \equiv \int [H(h'(x,y)) + c h'(x,y) + \mu (\rho^n - h'(x,y)^n)] \, dx \, dy,
    \end{align}
    for some Lagrange multiplier $\mu$. This means that it must satisfy
    \begin{align}
        H'(h_n(x,y)) + c - n \mu h_n(x,y)^{n-1} = 0,
    \end{align}
    with $\mu \ge 0$ and the complementary slackness condition
    \begin{align}
        \mu \left[ \rho^n - \int_{[0,1]^2} h_n(x,y)^n \, dx \, dy \right] = 0.
    \end{align}
    Note that we must have $\mu > 0$, since $H'(h_n) - c = 0$ would imply $h_n = \rho^*(c) \implies \lVert h_n \rVert_n = \rho^*(c)$, which violates the constraint. With $\mu > 0$, the function $H'(u) + c - n \mu u^{n-1}$ is strictly decreasing, and diverges as it approaches $0$ and $1$, so it must have a unique root. This means that there is a unique solution to the problem where the constraint binds, corresponding to $h_n = \rho$. Since our original graphon is in the feasible set of this problem, it is bounded by the value of the functional at the constant graphon. This yields the bound for the second case.
\end{proof}

Let us write ${\cal T}_2$ in terms of the decomposition of $u$:
\begin{align}
    {\cal T}_2(h,c) = \sum_{\theta \in \Theta} w_\theta \int_{[0,1]} \left[ \sum_{\theta' \in \Theta} c_{\theta \theta'} w_{\theta'} \int_{[0,1]} h_{\theta \theta'}(x,y) \, dy + \tilde{u}_\theta\left( \left[ \int_{[0,1]} h_{\theta \theta'}(x,y) \, dy \right]_{\theta' \in \Theta} \right) \right] \, dx.
\end{align}
In addition to this, we can also write the entropy term in terms of the subgraphons:
\begin{align}
    {\cal H}[h] = \sum_{\theta,\theta' \in \Theta} w_\theta w_{\theta'} \int_{[0,1]^2} H(h_{\theta,\theta'}(x,y)) \, dx \,dy.
\end{align}
Note that the Lemma \ref{lem:entropy_bound} allows us to jointly bound ${\cal T}_2 + {\cal H}$. Specifically, we obtain
\begin{align}
    {\cal T}_2(h,c) + {\cal H}(h) \le & \sum_{\theta,\theta' \in \Theta} w_\theta w_{\theta'} [H(\lVert h_{\theta \theta'} \rVert_{e^*}) + c_{\theta \theta'} \lVert h_{\theta \theta'} \rVert_{e^*}] \nonumber \\
    &+ \sum_{\theta \in \Theta} w_\theta \int_{[0,1]} \tilde{u}_\theta\left( \left[ \int_{[0,1]} h_{\theta \theta'}(x,y) \, dy \right]_{\theta' \in \Theta} \right) \, dx.
\end{align}
Using concavity of the $\tilde{u}_\theta$ functions, we can apply Jensen's inequality to obtain
\begin{align}
    \int_{[0,1]} \tilde{u}_\theta\left( \left[ \int_{[0,1]} h_{\theta \theta'}(x,y) \, dy \right]_{\theta' \in \Theta} \right) \le \tilde{u}_\theta\left( \left[ \int_{[0,1]^2} h_{\theta \theta'}(x,y) \, dx \, dy \right]_{\theta' \in \Theta} \right) = \tilde{u}_\theta\left( \left[ \lVert h_{\theta \theta'} \rVert_1 \right]_{\theta' \in \Theta} \right).
\end{align}
Since $\lVert h_{\theta \theta'} \rVert_1 \le \lVert h_{\theta \theta'} \rVert_{e^*}$, using monotonicity of $\tilde{u}_\theta$ yields
\begin{align}
    \int_{[0,1]} \tilde{u}_\theta\left( \left[ \int_{[0,1]} h_{\theta \theta'}(x,y) \, dy \right]_{\theta' \in \Theta} \right) \le \tilde{u}_\theta\left( \left[ \lVert h_{\theta \theta'} \rVert_{e^*} \right]_{\theta' \in \Theta} \right).
\end{align}
Together with the previous bound, this yields
\begin{align}
    {\cal T}_2(h,c) + {\cal H}(h) \le \sum_{\theta \in \Theta} w_\theta \left[ \sum_{\theta' \in \Theta} w_{\theta'} [H(\lVert h_{\theta \theta'} \rVert_{e^*}) + u_\theta([ \lVert h_{\theta \theta'} \rVert_{e^*} ]_{\theta' \in \Theta} ) \right].
\end{align}

As in the main text, let ${\cal K}_\Theta$ be the set of functions $\psi:\Theta^2 \to [0,1]$. For $\psi \in {\cal K}_\Theta$, define the function
\begin{align}
    Q(\psi) \equiv \sum_{m \in {\cal M}} a_m \sum_{\boldsymbol{\theta} \in \Theta^{n_m}} \left( \prod_{i \in {\cal N}_m} w_{\theta_i} \right) \prod_{ij \in m} \psi_{\theta_i \theta_j} + \sum_{\theta \in \Theta} w_\theta \left[ \sum_{\theta' \in \Theta} w_{\theta'} [H(\psi_{\theta \theta'}) + u_\theta([ \psi_{\theta \theta'} ]_{\theta' \in \Theta} ) \right].
\end{align}
Then the bounds above imply that, for an arbitrary colored graphon $(h,c)$,
\begin{align}
    {\cal T}(h,c) + {\cal H}(h) \le Q([\lVert h_{\theta \theta'} \rVert_{e^*}]_{\theta,\theta' \in \Theta}).
\end{align}
Define $\psi^*$ to be
\begin{align}
    \psi^* \in \argmax_{\psi \in {\cal K}_\Theta} \, Q(\psi).
\end{align}
This optimality implies that any graphon satisfies the following bound:
\begin{align}
    {\cal T}(h,c) + {\cal H}(h) \le Q(\psi^*).
\end{align}
Let $h^*$ be the piecewise constant graphon given by $h^*_{\theta \theta'}(x,y) = \psi^*_{\theta \theta'}$ for all $\theta, \theta', x, y$. Note that this graphon satisfies $\lVert h^*_{\theta \theta'} \rVert_n = \psi^*_{\theta \theta'}$ for all $n \ge 1$. In addition, all the inequalities used to construct the bounds above are equalities for this case, so we conclude that
\begin{align}
    {\cal T}(h^*,c) + {\cal H}(h^*) = Q(\psi^*).
\end{align}
Therefore, the graphon $h^*$ solves the variational problem in Theorem \ref{thm:colored_graphon_partition}. 

To show that only graphons that are constant almost everywhere solve the problem, assume for the sake of contradiction that there is a solution $\hat{h}$ that is not constant almost everywhere. Then the inequalities above are strict, and we would obtain a strict improvement by ``flattening'' the solution in a way that preserves the norms of the subgraphons $\lVert \hat{h}_{\theta \theta'} \rVert_{e^*}$.

The convergence to the directed stochastic block model when the optimizer is unique is a direct consequence of Theorem \ref{thm:colored_graphon_convergence}.

\subsection{Proofs of Lemmas}
\subsubsection{Proof of Lemma \ref{lem:forward_looking}}
Note that the value functions $V_i^\Phi$ satisfy Equation \eqref{eq:hjb}, meaning they are a fixed point of the map $T$ used in the proof of Proposition \ref{prop:MPE}. Therefore, they correspond to Markov-perfect equilibria of the game.

\subsubsection{Proof of Lemma \ref{lem:fixed_point}}
Due to the divergence of $H'(u)$ near $u = 0$ and $u = 1$, we have that any solution $\rho^*$ to the problem in Proposition \ref{prop:conv} must be interior. Therefore, it must satisfy the first-order condition
\begin{align} \label{eq:motif_foc}
    \sum_{m \in {\cal M}} a_m e_m (\rho^*)^{e_m - 1} + H'(\rho^*) = 0.
\end{align}
This can be written as
\begin{align}
    K_{{\cal M}}(\rho^*) = \log\left( \frac{\rho^*}{1-\rho^*} \right).
\end{align}
Inverting the function to the right gives
\begin{align*}
    \rho^* = R(\rho^*).
\end{align*}

\subsubsection{Proof of Lemma \ref{lem:nash_equilibria}}
First consider the case $a_{m_1} < - a_{m_2}$. This means that the ``cost'' of establishing a directed link is larger than the benefit of a mutual link. This means that if agent $i$ has a link towards agent $j$, it will always be beneficial to deviate by removing the link, regardless of whether agent $j$ also has a link towards agent $i$. Therefore, the only equilibrium is the empty network.

The second case is when the cost of establishing a directed link is lower than the benefit of a mutual link. Networks that are Nash equilibria are those for which no agent can strictly increase their payoff by changing their outward connections. For a given network, an agent's best response is to match connections that are directed towards them and place no additional connections. Therefore, only networks where all links are reciprocated will be Nash equilibria of the deterministic game.

\subsubsection{Proof of Lemma \ref{lem:trade_fixed_point}}
Similar to the proof of Lemma \ref{lem:fixed_point}, the divergence in $H'(u)$ near 0 and 1 ensures the solution to the problem in Equation \eqref{eq:trade_partition} is interior. For $\theta' \ne \theta$, the first-order condition for $\psi_{\theta \theta'}$ is
\begin{align}
    w_\theta w_{\theta'} \left[ \frac{r}{2} (\psi^*_{\theta \theta'} + \psi^*_{\theta' \theta}) + H'(\psi^*_{\theta \theta'}) - \gamma D(\theta,\theta') \right] = 0.
\end{align}
Taking the first-order condition for $\psi_{\theta' \theta}$ and subtracting the equation above yields
\begin{align}
    H'(\psi^*_{\theta \theta'}) = H'(\psi^*_{\theta' \theta}) \implies \psi^*_{\theta \theta'} = \psi^*_{\theta' \theta},
\end{align}
since $D(\theta, \theta') = D(\theta', \theta)$. Therefore, $\psi^*_{\theta \theta'}$ satisfies
\begin{align}
    (r \psi^*_{\theta \theta'} - \gamma D(\theta, \theta')) + H'(\psi^*_{\theta \theta'}) = 0.
\end{align}
Note that this is the same as Equation \eqref{eq:motif_foc} with values ${\bf a} = (- \gamma D(\theta,\theta'), r/2)$.

For $\theta = \theta'$, the first-order condition is
\begin{align}
    w_\theta^2\left[ (r \psi^*_{\theta \theta} - \gamma D(\theta,\theta)) + H'(\psi^*_{\theta \theta}) \right] = 0.
\end{align}
Therefore, the same fixed-point problem is solved by all pairs $(\theta, \theta')$ with their respective distances.

\subsubsection{Proof of Lemma \ref{lem:trade_global_fixed_point}}
Since ${\bf w}$ is uniform, we have $w_\theta = 1/L$ for all $\theta$. For a chain motif $c_\ell$ with $\ell$ links, its motif density is
\begin{align}
    b[c_\ell, \psi ; {\bf w}] = \frac{1}{L^{\ell+1}} \sum_{\boldsymbol{\theta} \in \Theta^{\ell+1}} \left( \prod_{i=1}^{\ell} \psi_{\theta_i \theta_{i+1}} \right).
\end{align}
The partition function optimization problem for this model is
\begin{align}
    \zeta(\gamma,r) = \max_{\psi \in {\cal K}_\Theta} \left[ \frac{r}{L^{\ell+1}} \sum_{\boldsymbol{\theta} \in \Theta^{\ell+1}} \left( \prod_{i=1}^{\ell} \psi_{\theta_i \theta_{i+1}} \right) + \frac{1}{L^2} \sum_{\theta,\theta' \in \Theta^2} \left( H(\psi_{\theta \theta'}) -\gamma \psi_{\theta \theta'} D(\theta,\theta') \right) \right].
\end{align}

By the same argument as in the previous lemmas, the solution must be interior. Therefore, it must satisfy the first-order condition
\begin{align}
    \frac{r}{L^{\ell+1}} \sum_{i = 1}^\ell \sum_{\boldsymbol{\theta} \in \Theta^{\ell+1}} \left( \prod_{\substack{1 \le j \le \ell \\ j \ne i}} \psi^*_{\theta_i \theta_{i+1}} \right) \mathbbm{1}\{\theta_i = \theta, \theta_{i+1} = \theta'\} + \frac{1}{L^2} \left( H'(\psi^*_{\theta \theta'}) -\gamma D(\theta,\theta') \right) = 0.
\end{align}
Note that if the maximizer is unique, then $\psi^*_{\theta \theta'}$ can only depend on $D(\theta,\theta')$. This is because if it were not the case, then we could create a different kernel by shifting $\psi^*$ clockwise one unit, and this kernel would yield the same value of the objective function, violating uniqueness. This implies that the sum
\begin{align}
    \frac{1}{L} \sum_{\theta_2 \in \Theta} \psi^*_{\theta_1 \theta_2}
\end{align}
is the same for all $\theta_1$. Let us denote this sum with $\rho^*$. In the first-order condition above, each term indexed by $i$ fixes the coordinates $i$ and $i+1$, and sums over the rest. The sums over $\Theta^{\ell+1}$ can be done iteratively from the highest $j$, and they will all yield the value $\rho^*$. Therefore, the first-order condition takes the form
\begin{align}
    \frac{1}{L^2} ( \ell r (\rho^*)^{\ell-1} + H'(\psi^*_{\theta \theta'}) - \gamma D(\theta, \theta')) = 0.
\end{align}
Inverting this equation yields
\begin{align}
    \psi^*_{\theta \theta'} = \left[ 1 + \exp(\gamma D(\theta,\theta') - \ell r (\rho^*)^{\ell - 1} ) \right]^{-1}.
\end{align}

Now, with this we can compute the sum that determines $\rho^*$. Fixing $\theta$, we obtain
\begin{align}
    \rho^* = \frac{1}{L} \sum_{\theta' \in \Theta} \left[ 1 + \exp(\gamma D(\theta,\theta') - \ell r (\rho^*)^{\ell - 1} ) \right]^{-1}.
\end{align}
Since this is independent of $\theta$, it can also be written as an average over $\theta$:
\begin{align}
    \rho^* = \frac{1}{L^2} \sum_{\theta,\theta' \in \Theta} \left[ 1 + \exp(\gamma D(\theta,\theta') - \ell r (\rho^*)^{\ell - 1} ) \right]^{-1},
\end{align}
thus yielding the result.
\section{Graph Limits} \label{sec:app_graph_limits}
In this section, I present the relevant definitions and results for the theory of large dense graphs. The results in Sections \ref{subsec:graphons} and \ref{subsec:results_graph_limits} are presented without proof. These are taken from Appendix D in \citet{Mele2017AFormation}, which is an excellent introduction to the topic, and builds on the results in \citet{chatterjee2011large} and \citet{Chatterjee2013}. The reader is invited to read this, and the references therein, for a detailed discussion on the topic\footnote{Also see \citet{lovasz2012large} for a broader introduction to graph limits.}. Section \ref{subsec:colored_graphons} extends these results to colored graphs, allowing us to characterize the limiting behavior of the model with heterogeneous agents. In order to make a clearer connection to the literature on graph limits, the notation in this section will differ from the rest of the paper. 

\subsection{Graphons} \label{subsec:graphons}
This section presents the relevant definitions for the theory of graph limits, following the outline presented in \citet{Chatterjee2013}. For this appendix, a directed graph is an ordered pair $G = (V,A)$, where $V$ is a set of vertices and $A$ is a set of arcs (directed edges). Consider a sequence $G_N$ of simple directed graphs whose number of nodes tends to infinity. For every fixed simple graph $H$, let $|\textrm{hom}(H,G)|$ denote the number of homomorphisms of $H$ into $G$, which is the number of edge-preserving maps from $V(H)$ into $V(H)$. That is, a map $\varphi:V(H) \to V(G)$ is a homomorphism if $(i,j)\in A(H) \implies (\varphi(i),\varphi(j)) \in A(G)$. Normalizing by the possible number of maps from $V(H)$ to $V(H)$ yields the \textit{homomorphism density}
\begin{align}
    t(H,G) \equiv \frac{|\textrm{hom}(H,G)|}{|V(G)|^{|V(H)|}},
\end{align}
which corresponds to the probability of a uniformly random mapping $V(H) \to V(H)$ is a homomorphism.

The importance of analyzing homomorphism densities is twofold. First, they provide a way to ``probe'' large graphs in order to understand their properties. Second, they're relevant in our context since motif utilities can be approximated as homomorphism densities for large graphs, as is discussed below.

Motif utilities do not depend explicitly on the number of homomorphisms of a motif into the network, but rather the number of times it appears as a subgraph. This can be captured using the concept of \textit{subgraph densities}. Let $|\textrm{sub}(H,G)|$ be the number of \textit{injective} maps from $V(H)$ to $V(G)$ that are homomorphisms. Clearly $|\textrm{sub}(H,G)| \le |\textrm{hom}(H,G)|$. The normalized motif densities in our model are given by subgraph densities, defined by
\begin{align}
    s(H,G) \equiv \frac{|\textrm{sub}(H,G)|}{|V(G)|^{|V(H)|}}.
\end{align}
It can be shown that
\begin{align}
    t(H,G) - \frac{1}{|V(G)|} {|V(H)| \choose 2} \le s(H,G) \le t(H,G).
\end{align}
Therefore, for a given $H$, characterizing homomorphism densities for large graphs is equivalent to characterizing subgraph densities.

A possible notion of convergence for the sequence $\{G_N\}$ is that the densities $t(H,G_N)$ converge to some value for every finite graph $H$. The work of Lov\'asz and coauthors (see \citet{lovasz2012large} for an overview) established the existence of an object that characterizes this convergence for undirected graphs. That is, the exists an object from which the limiting homomorphism densities $t(H,\cdot)$ can be obtained. An analogous result for directed graphs was established in \citet{boeckner2013directed}. This ``limiting object'' is a function $h \in {\cal W}$, where ${\cal W}$ is the set of measurable functions $[0,1]^2 \to [0,1]$. These objects are called ``graphons''. Conversely, every function in ${\cal W}$ arises as the limit of an appropriate sequence of directed graphs. If $H$ is a simple directed graph with $V(H) = [k] = \{1,\ldots,k\}$, the homomorphism density of $H$ into $h$ is defined as
\begin{align}
    t(H,h) \equiv \int_{[0,1]^k} \left( \prod_{(i,j) \in E(H)} h(x_i, x_j) \right) \, dx_1 \ldots dx_k.
\end{align}
A sequence of graphs $\{G_N\}$ is said to converge to $h$ if 
\begin{align}
    \lim_{N \to \infty} t(H,G_N) = t(H,h)
\end{align}
for all finite simple directed graphs $H$.

The intuition is that as $N \to \infty$, the interval $[0,1]$ represents a ``continuum'' of vertices and $h(x,y)$ is the probability that there is an arc going from $x$ to $y$.\footnote{This intuition is more explicit in the case of $W$ random graphs.} For the case of directed Erd\"os-R\'enyi graphs $G(N,p)$ with fixed $p$, the limit graph is represented by the graphon that is equal to $p$ for all $(x,y) \in [0,1]^2$. For a fixed $p \in (0,1)$, this corresponds to a dense random graph model.

Finite simple graphs have a canonical representation as graphons. For a given graph $G$ over $N$ nodes, its associated graphon is given by
\begin{align}
    h^G(x,y) \equiv \mathbbm{1}\{ (\lceil Nx \rceil, \lceil Ny \rceil) \in A(G) \}.
\end{align}

The notion of convergence in ${\cal W}$ in terms of homomorphism densities can be metrized using the \textit{cut distance}. For two graphons $h_1$ and $h_2$, it is defined as
\begin{align}
    d_\square(h_1,h_2) \equiv \sup_{S, T \subseteq [0,1]} \left| \int_{S \times T} [h_1(x,y) - h_2(x,y)] \, dx \, dy \right|.
\end{align}
For our purposes, it will be useful to work in a different space that has some useful topological properties. Let $\Sigma$ be the set of measure-preserving bijections $\sigma: [0,1] \to [0,1]$. Define an equivalence relation on ${\cal W}$ by setting $h_1 \sim h_2$ if $h_1(x,y) = h_2^\sigma(x,y) \equiv h_2(\sigma x, \sigma y)$ for some $\sigma \in \Sigma$. Let $\tilde{h}$ be the closure of the orbit $\{h^\sigma\}$ in $({\cal W},d_\square)$. Let $\tilde{{\cal W}} \equiv {\cal W}/\sim$ and let $\tau$ be the map $\tau h \mapsto \tilde{h}$. The distance $\delta_\square$ on the space $\tilde{{\cal W}}$ is defined as
\begin{align}
    \delta_\square(\tilde{h}_1, \tilde{h}_2) = \inf_{\sigma} d_\square(h_1, h_2^\sigma),
\end{align}
such that $(\tilde{{\cal W}}, \delta_\square)$ is a metric space. For any finite directed graph $G$, let $G$ be its associated graphon and let $\tilde{G} \equiv \tilde{h}^G \in \tilde{{\cal W}}$ be its corresponding orbit.

An important aspect of homomorphism densities is that they are continuous functions in this space. This is useful in the analysis of the limiting behavior of ERGMs, as discussed below.

\subsection{Results on graph limits} \label{subsec:results_graph_limits}
A central tool used in characterizing the limiting behavior of ERGMs is the theory of large deviations for random graph models. Fix some $p \in (0,1)$. Let $\tilde{\mathbb{P}}_{N,p}$ be the measure induced on $\tilde{{\cal W}}$ by the directed Erd\"os-R\'enyi model with parameter $p$. Additionally, define the function $I_p:[0,1] \to \mathbb{R}$ as
\begin{align}
    I_p(u) \equiv u \log\left( \frac{u}{p} \right) + (1-u) \log\left( \frac{1-u}{1-p} \right).
\end{align}
The domain of this function can be extended to $\tilde{{\cal W}}$ by defining
\begin{align}
    {\cal I}_p(\tilde{h}) \equiv \int_{[0,1]^2} I_p(h(x,y)) \, dx \, dy
\end{align}
for any $h \in \tau^{-1}(\tilde{h})$. Building off the results in \citet{chatterjee2011large}, we have the following result for directed graphs.
\begin{theorem}[Theorem 8 in \citet{Mele2017AFormation}]\label{thm:er_ldp}
    For each fixed $p \in (0,1)$, the sequence $\tilde{\mathbb{P}}_{N,p}$ obeys a large deviation principle on the space $(\tilde{W}, \delta_\square)$ with rate function $\tilde{I}_p$. Explicitly, for any closed set $\tilde{F} \subseteq \tilde{{\cal W}}(\tilde{F})$,
    \begin{align}
        \limsup_{N \to \infty} \frac{1}{N^2} \log(\tilde{\mathbb{P}}_{N,p}(\tilde{F})) \le - \inf_{\tilde{h} \in \tilde{F}} {\cal I}_p(\tilde{h})
    \end{align}
    and for any open set $\tilde{U} \subseteq \tilde{{\cal W}}$,
    \begin{align}
        \limsup_{N \to \infty} \frac{1}{N^2} \log(\tilde{\mathbb{P}}_{N,p}(\tilde{U})) \ge - \inf_{\tilde{h} \in \tilde{U}} {\cal I}_p(\tilde{h}).
    \end{align}
\end{theorem}

Now suppose we have a sequence of measures $\pi_N$ on ${\cal G}_N$ given by
\begin{align}
    \pi_N(G) = \exp\{ N^2 ({\cal T}(\tilde{G}) - \psi_N) \},
\end{align}
where $\psi_N$ is a normalization constant given by
\begin{align}
    \psi_N \equiv \frac{1}{N^2} \log\left( \sum_{G \in {\cal G}} e^{N^2 {\cal T}(\tilde{G})} \right).
\end{align}
In the case of \citet{Mele2017AFormation} and our motif utilities, the function ${\cal T}$ is a linear combination of homomorphism densities. Define the \textit{entropy functional} ${\cal H}:\tilde{{\cal W}} \to \mathbb{R}$
\begin{align}
    {\cal H}(\tilde{h}) = \int_{[0,1]^2} H(h(x,y)) \, dx \, dy
\end{align}
for some $h \in \tau^{-1}(\tilde{h})$ and $H$ is given by
\begin{align}
    H(u) \equiv - u \log(u) - (1-u) \log(1-u).
\end{align}
Adapting the results in \citet{Chatterjee2013}, \citet{Mele2017AFormation} finds the following result on the normalization constant of these models.

\begin{theorem}[Theorem 10 in \citet{Mele2017AFormation}]\label{thm:partition_func_convergence}
    If ${\cal T}:\Tilde{{\cal W}} \to \mathbb{R}$ is a bounded continuous function, then
    \begin{align} \label{eq:partition_func_convergence}
        \psi \equiv \lim_{N \to \infty} \psi_N = \sup_{\Tilde{h} \in \Tilde{{\cal W}}} \left[ {\cal T}(\Tilde{h}) + {\cal H}(\Tilde{h}) \right].
    \end{align}
\end{theorem}

Using the properties of the space $\tilde{{\cal W}}$, an even stronger result on convergence can be obtained. This allows us to characterize the limiting behavior of the statistical properties of ERGMs.

\begin{theorem}[Theorem 18 in \citet{Mele2017AFormation}]\label{thm:graphon_convergence}
    Let $\Tilde{M}^*$ be the set of maximizers of the variational problem \eqref{eq:partition_func_convergence}. Let $G_N$ be a graph on $N$ vertices drawn from the model implied by function ${\cal T}$. Then, for any $\eta > 0$, there exist $C, \kappa > 0$ such that, for any $N$,
    \begin{align*}
        \mathbb{P}\{ \delta_\square(\Tilde{G}_N, \Tilde{M}^*) > \eta \} \le C e^{-N^2 \kappa},
    \end{align*}
    where $\mathbb{P}$ denotes the probability measure implied by the model.
\end{theorem}

Building on these results, we can find characterize the limiting behavior of models with heterogeneity in the nodes.

\subsection{Colored graphons} \label{subsec:colored_graphons}
In order to account for node heterogeneity, I use the framework of \textit{colored graphs}, particularly as laid out in \citet{diao2016model}. Let $\Theta$ be a set of colors (or types in our case). A directed colored graph $Q$ is a tuple $(V,A,C)$, where $C:V \to \Theta$ represents the coloring of the vertices of $Q$.

Similar to the case of uncolored graphs, we can consider a way to ``probe'' colored graphs using the homomorphism densities of other colored graphs into them. Let $R$ and $Q$ be directed colored graphs. We define $|\textrm{hom}_\Theta(R,Q)|$ to be the number of homomorphisms of $(V(R),A(R))$ into $(V(Q),A(Q))$ such that the corresponding map preserves the coloring of the vertices. The \textit{colored homomorphism density} of $R$ into $Q$, then, is
\begin{align}
    t_\Theta(R,Q) \equiv \frac{|\textrm{hom}_\Theta(R,Q)|}{|V(Q)|^{|V(R)|}}.
\end{align}
Like in the uncolored case, we can characterize the convergence of a sequence of a series of colored graphs by studying their colored homomorphism densities. In order to do this, we need to define appropriate limiting objects.

A colored directed graphon is a pair $q \equiv (h_q, c_q)$, where $h_q \in {\cal W}$ is a directed graphon and $c_q$ is a measurable function $c: [0,1] \to \Theta$. The space of colored directed graphons is denoted with ${\cal W}_\Theta$. The colored homomorphism density of $R$ into $q$, assuming $V(R) = [k]$, is defined as
\begin{align}
    t_\Theta(R,q) \equiv \int_{[0,1]^{k}} \left( \prod_{(i,j) \in A(R)} h_q(x_i,x_j) \right) \left( \prod_{i=1}^k \mathbbm{1}\{ c_q(x_i) = C_R(i) \} \right) \, dx_1 \ldots dx_{k}.
\end{align}
As in the uncolored case, there is a canonical representation of colored graphs as colored graphons. For a given colored graph $Q$, its associated graphon $q^Q = (h_{q^Q}, c_{q^Q})$ is given by
\begin{align}
    h_{q^Q} = h^{(V(Q),A(Q))}, \quad c_{q^Q} = \sum_{\theta \in \Theta} \theta \mathbbm{1}\{ C_Q(\lceil |V(Q)| x \rceil) = \theta \}.
\end{align}

We can also construct a topology in the space of colored graphons. In order to do this, we need a new notion of distance. The \textit{colored cut distance} between two graphons is given by
\begin{align}
    d_\square^\Theta(q_1, q_2) \equiv d_\square(h_{q_1}, h_{q_2}) + d_\Theta(c_{q_1},c_{q_2}),
\end{align}
where
\begin{align}
    d_\Theta(c, c') \equiv \sum_{\theta \in \Theta} \int_{[0,1]} \mathbbm{1}\{ x \in c^{-1}(\theta) \Delta c'^{-1}(\theta) \} \, dx
\end{align}
is a distance on the space of colorings. The equivalence class in this space is defined similarly to the uncolored case, but we have to take into account that the bijections $\sigma$ also act on the colorings. Define the equivalence relation $\sim$ by setting $q_1 \sim q_2$ if $h_{q_1}(x,y) = h_{q_2^\sigma}(x,y) = h_{q_2}(\sigma x, \sigma y)$ and $c_{q_1}(x) = c_{q_2^\sigma}(x) = c_{q_2}(\sigma x)$ for some $\sigma \in \Sigma$. Define $\tilde{q}$ to be the equivalence class of $q$, and let $\tilde{{\cal W}}_\Theta \equiv {\cal W}_\Theta/\sim$. Additionally, let $\tau_\Theta$ be the map such that $\tau_\Theta(q) = \tilde{q}$. The natural distance on this space is given by
\begin{align}
    \delta_\square^\Theta(\tilde{q}_1, \tilde{q}_2) \equiv \inf_\sigma d_\square^\Theta(q_1, q_2^\sigma).
\end{align}
It will also be convenient to define an analogous equivalence on the space of colorings, and to define the coloring distance
\begin{align}
    \delta_\Theta(\tilde{c},\tilde{c}') = \inf_\sigma d_\Theta(c,c'^\sigma).
\end{align}
The space $(\tilde{{\cal W}}_\Theta, \delta^\Theta_\square)$ has similar topological properties to $(\tilde{{\cal W}}, \delta_\square)$. There is one particular result that is of interest to us, since it allows us to characterize the limiting behavior of the measures induced by random graph models.

\begin{theorem}
    The space $(\tilde{{\cal W}}_\Theta, \delta_\square^\Theta)$ is compact.
\end{theorem}

\begin{proof}
    This proof follows similarly to the proof of Theorem 3.7 in \citet{diao2016model}, with suitable regularity conditions for the directed graph case used in the proof of Lemma 5 in \citet{Mele2017AFormation}.
\end{proof}

Having this topological property, we can now state the result that allows for the generalization of the graph convergence results to the colored case. Consider a sequence of empirical fractions of colors ${\bf w}^N$ that converges to some ${\bf w}$. Let $\Tilde{\mathbb{P}}^\Theta_{N,p}$ be the sequence of measures on $\tilde{{\cal W}}_\Theta$ induced by having graphs with colorings with empirical fractions ${\bf w}^N$ and sampling their edges using an Erd\"os-R\'enyi model with parameter $p$. The following is a slightly different statement of the large deviations principle, but it is equivalent to the one in Theorem \ref{thm:er_ldp}.
\begin{theorem}
    For each fixed $p \in (0,1)$, the sequence $\Tilde{\mathbb{P}}^\Theta_{N,p}$ obeys a large deviation principle in the space ($\Tilde{{\cal W}}_\Theta$, $\delta_{\square}^\Theta$) with rate function
    \begin{align}
        {\cal J}^\Theta_{{\bf w},p}(\tilde{f}) = 
        \begin{cases}
            {\cal I}_p(\tilde{h}_f) & \textrm{if } \delta_\Theta(\tilde{c}_f,\tilde{c}_{{\bf w}}) = 0, \\
            \infty & \textrm{otherwise}.
        \end{cases}
    \end{align}
    Explicitly, this means that for any Borel set $\tilde{E} \subseteq \tilde{{\cal W}}_\Theta$,
    \begin{align}
        - \inf_{\tilde{f} \in \textrm{int}(\tilde{E})} {\cal J}^\Theta_{{\bf w},p}(\tilde{f})\le \liminf_{N \to \infty} \frac{1}{N^2} \log(\tilde{\mathbb{P}}^\Theta_{N,p}(\tilde{E})) \le \limsup_{N \to \infty} \frac{1}{N^2} \log(\tilde{\mathbb{P}}^\Theta_{N,p}(\tilde{E})) \le - \inf_{\tilde{f} \in \textrm{cl}(\tilde{E})} {\cal J}^\Theta_{{\bf w},p}(\tilde{f}).
    \end{align}
\end{theorem}

\begin{proof}
    Let $\tilde{E} \subseteq \tilde{{\cal W}}$ be a Borel set and let $\tilde{E}_N$ be the (discrete) set of colored graphons that can be obtained with the empirical fractions ${\bf w}^N$ and are a subset of $\tilde{E}$. We can see that
    \begin{align}
        \Tilde{\mathbb{P}}^\Theta_{N,p}(\tilde{E}) = \Tilde{\mathbb{P}}^\Theta_{N,p}(\tilde{E}_N).
    \end{align}
    Define the function $K:\tilde{{\cal W}}_\Theta \to {\cal W}$ by
    \begin{align}
        K(\tilde{q}) \equiv \tilde{h}_f.
    \end{align}
    This function acts as a ``projection'' onto the space of graphons. This projection will allow us to relate the measure $\tilde{\mathbb{P}}^\Theta_{N,p}$ to the Erd\"os-R\'enyi measure on $\tilde{{\cal W}}$. First, note that
    \begin{align}
        \Tilde{\mathbb{P}}^\Theta_{N,p}(\tilde{E}_N) \le \Tilde{\mathbb{P}}_{N,p}(K(\tilde{E}_N)),
    \end{align}
    since the mass given to every feasible graphon in $\tilde{{\cal W}}$ is distributed among the corresponding colored graphons. We can also lower bound this probability. Let $\phi$ be a permutation of $[N]$. For a finite graph $G$ with $N$ nodes, the probability associated to $\tilde{G}$ is
    \begin{align}
        \tilde{\mathbb{P}}_{N,p}(\{\tilde{G}\}) = p^{|A(G)|} (1-p)^{N(N-1)-|A(G)|} N! \left( \sum_{\phi} \mathbbm{1}\{\phi A(G) = A(G)\} \right)^{-1},
    \end{align}
    where the last factor takes into account the overcounting of permutations. Similarly, for a colored graph $Q$ over $N$ nodes, its associated probability is
    \begin{align}
        \tilde{\mathbb{P}}^\Theta_{N,p}(\{\tilde{Q}\}) = p^{|A(Q)|} (1-p)^{N(N-1)-|A(Q)|} \left( \prod_{\theta \in \Theta} (N w_\theta^N)! \right) \left( \sum_{\phi} \mathbbm{1}\{\phi A(Q) = A(Q), \phi C_Q = C_Q\} \right)^{-1},
    \end{align}
    where the overcounting correction now also takes into account the overcounting of colors. From these two expressions, we can see that if $G = (V(Q),A(Q))$, then
    \begin{align}
        \tilde{\mathbb{P}}^\Theta_{N,p}(\{\tilde{Q}\}) \ge \frac{\left( \prod_{\theta \in \Theta} (N w_\theta^N)! \right)}{N!} \tilde{\mathbb{P}}_{N,p}(\{\tilde{G}\}).
    \end{align}
    Since these bounds hold for all graphs, we have that 
    \begin{align}
        \frac{\left( \prod_{\theta \in \Theta} (N w_\theta^N)! \right)}{N!} \tilde{\mathbb{P}}_{N,p}(K(\tilde{E}_N)) \le \Tilde{\mathbb{P}}^\Theta_{N,p}(\tilde{E}) \le \Tilde{\mathbb{P}}_{N,p}(K(\tilde{E}_N)).
    \end{align}
    Note that the factorial terms vanish compared to the $N^2$ scaling that we need to consider for the large deviations principle. Explicitly, we have that
    \begin{align}
        \liminf_{N \to \infty} \frac{1}{N^2} \log(\Tilde{\mathbb{P}}^\Theta_{N,p}(\tilde{E})) &= \liminf_{N \to \infty} \frac{1}{N^2} \log(\Tilde{\mathbb{P}}_{N,p}(K(\tilde{E}_N))), \nonumber \\
        \limsup_{N \to \infty} \frac{1}{N^2} \log(\Tilde{\mathbb{P}}^\Theta_{N,p}(\tilde{E})) &= \limsup_{N \to \infty} \frac{1}{N^2} \log( \Tilde{\mathbb{P}}_{N,p}(K(\tilde{E}_N))).
    \end{align}
    Since these limits are in terms of the Erd\"os-R\'enyi measures on $\tilde{{\cal W}}$, we can now manipulate these expressions to relate them to their corresponding large deviation principle.

    Fix $\epsilon > 0$. Define
    \begin{align}
        \tilde{D}_{\textrm{w}} \equiv \{ \tilde{f} \in \tilde{{\cal W}}_\Theta : \delta _\Theta(\tilde{c}_f,\tilde{c}_{\bf w}) = 0 \},
    \end{align}
    where $\tilde{c}_{\bf w})$ is the equivalence class of a coloring consistent with ${\bf w}$. Since ${\bf w}^N \to {\bf w}$, we have that $\delta_\Theta(\tilde{c}_{{\bf w}^N}, \tilde{c}_{{\bf w}}) \to 0$. Additionally, define the set
    \begin{align}
        \tilde{D}_{{\bf w}}^\epsilon \equiv \{ \tilde{q} \in \tilde{{\cal W}}_\Theta : \delta_\Theta(\tilde{c}_q, \tilde{c}_{{\bf w}}) \}.
    \end{align}
    Note that for $N$ large enough, $\tilde{E}_N \subseteq \tilde{E} \cap \tilde{D}_{{\bf w}}^\epsilon$. Now, we have that if $\tilde{A} \subseteq \tilde{B}$, then $K(\tilde{A}) \subseteq K(\tilde{B})$. Therefore, we have that
    \begin{align}
        \limsup_{N \to \infty} \frac{1}{N^2} \log( \Tilde{\mathbb{P}}_{N,p}(K(\tilde{E}_N))) \le \limsup_{N \to \infty} \frac{1}{N^2} \log( \Tilde{\mathbb{P}}_{N,p}(K(\tilde{E} \cap \tilde{D}_{{\bf w}}^\epsilon))).
    \end{align}
    Using our previous results and Theorem \ref{thm:er_ldp}, this yields
    \begin{align}
        \limsup_{N \to \infty} \frac{1}{N^2} \log(\Tilde{\mathbb{P}}^\Theta_{N,p}(\tilde{E})) \le - \inf_{\tilde{h} \in \textrm{cl}(K(\tilde{E} \cap \tilde{D}_{{\bf w}}^\epsilon))} {\cal I}_p(\tilde{h}).
    \end{align}
    
    To obtain an analogous lower bound, consider the set
    \begin{align}
        \tilde{E}^\epsilon \equiv \left\{ \tilde{q} \in \tilde{{\cal W}}_\Theta : \inf_{\tilde{q}' \in \tilde{E}^c} \delta_\square^\Theta(\tilde{q}, \tilde{q}') > \epsilon \right\}.
    \end{align}
    Now, consider a colored graphon $\tilde{q}_0 \in \tilde{E}^\epsilon \cap \tilde{D}_{{\bf w}}$ such that $\tilde{h}_{q_0}$ is a feasible graphon over $N$ nodes. I will show that, for $N$ sufficiently large, $\tilde{h}_{q_0}$ must also correspond to some colored graphon in $\tilde{E}_N$. Suppose this is not the case. Consider the colored graphon $q_1 = (h_{q_0}, c^*)$, where
    \begin{align}
        c^* \in \argmin_c d_\Theta(c,c_{q_0}) \quad \textrm{s.t. } \tilde{c} = \tilde{c}_{{\bf w}^N}. 
    \end{align}
    For $N$ sufficiently large, we will have $d_\Theta(c^*,c_{q_0}) < \epsilon$. This means that $d_\square^\Theta(q_0, q_1) < \epsilon$, so $\delta_\square^\Theta(\tilde{q}_0, \tilde{q}_1) < \epsilon$. Given the definition of $\tilde{E}^\epsilon$, this implies that $\tilde{q}_1 \in \tilde{E}$, which is a contradiction. This means that, for sufficiently large $N$,
    \begin{align}
        \tilde{\mathbb{P}}_{n,q}(K(\tilde{E}^\epsilon \cap \tilde{D}_{{\bf w}})) \le \tilde{\mathbb{P}}_{n,q}(K(\tilde{E}_N)).
    \end{align}
    Therefore,
    \begin{align}
        \liminf_{N \to \infty} \frac{1}{N^2} \log(\Tilde{\mathbb{P}}_{N,p}(K(\tilde{E}^\epsilon \cap \tilde{D}_{{\bf w}}))) \le \liminf_{N \to \infty} \frac{1}{N^2} \log(\Tilde{\mathbb{P}}_{N,p}(K(\tilde{E}_N))).
    \end{align}
    Again using the large deviations result from Theorem \ref{thm:er_ldp}, this yields
    \begin{align}
        - \inf_{\tilde{h} \in \textrm{int}(K(\tilde{E}^\epsilon \cap \tilde{D}_{{\bf w}}))} {\cal I}_p(\tilde{h}) \le \liminf_{N \to \infty} \frac{1}{N^2} \log(\Tilde{\mathbb{P}}^\Theta_{N,p}(\tilde{E})).
    \end{align}
    Putting these results together, we have
    \begin{align}
        - \inf_{\tilde{h} \in \textrm{int}(K(\tilde{E}^\epsilon \cap \tilde{D}_{{\bf w}}))} {\cal I}_p(\tilde{h}) &\le \liminf_{N \to \infty} \frac{1}{N^2} \log(\Tilde{\mathbb{P}}^\Theta_{N,p}(\tilde{E})) \nonumber \\
        &\le \limsup_{N \to \infty} \frac{1}{N^2} \log(\Tilde{\mathbb{P}}^\Theta_{N,p}(\tilde{E})) \le - \inf_{\tilde{h} \in \textrm{cl}(K(\tilde{E} \cap \tilde{D}_{{\bf w}}^\epsilon))} {\cal I}_p(\tilde{h}),
    \end{align}
    which holds for all $\epsilon > 0$. Taking $\epsilon \to 0$, we obtain
    \begin{align}
        - \inf_{\tilde{h} \in \textrm{int}(K(\tilde{E} \cap \tilde{D}_{{\bf w}}))} {\cal I}_p(\tilde{h}) &\le \liminf_{N \to \infty} \frac{1}{N^2} \log(\Tilde{\mathbb{P}}^\Theta_{N,p}(\tilde{E})) \nonumber \\
        &\le \limsup_{N \to \infty} \frac{1}{N^2} \log(\Tilde{\mathbb{P}}^\Theta_{N,p}(\tilde{E})) \le - \inf_{\tilde{h} \in \textrm{cl}(K(\tilde{E} \cap \tilde{D}_{{\bf w}}))} {\cal I}_p(\tilde{h}).
    \end{align}
    This almost has the form of the large deviations result. We need to show that the bounds can be written in terms of infima over subsets of $\tilde{{\cal W}}_\Theta$ instead of $\tilde{{\cal W}}$. Note that
    \begin{align}
        \inf_{\tilde{q} \in \tilde{E}} {\cal J}^\Theta_{p,{\bf w}}(\tilde{q}) = \inf_{\tilde{h} \in K(\tilde{E} \cap \tilde{D}_{{\bf w}})} {\cal I}_p(\tilde{h}).
    \end{align}
    This gives us a way to translate the bounds above into bounds arising in the space of colored graphons. 
    
    Now consider an open set $\tilde{U} \subseteq \tilde{{\cal W}}_\Theta$. I will prove that the set $K(\tilde{U} \cap \tilde{D}_{{\bf w}})$ is open. Suppose it is not. Then, there exists an $\tilde{h}_0 \in K(\tilde{U} \cap \tilde{D}_{{\bf w}})$ such that for all $\varepsilon > 0$, there is some $\tilde{h}_\varepsilon$ such that $\delta_\square(\tilde{h}_0,\tilde{h}_\varepsilon) < \varepsilon$ and $\tilde{h}_\varepsilon \notin K(\tilde{U} \cap \tilde{D}_{{\bf w}})$. Let us construct a colored graphon as follows: take some $h_0 \in \tau^{-1}(\tilde{h}_0)$. Additionally, take $h_\varepsilon \in \tau^{-1}(\tilde{h}_\varepsilon)$ such that $d_\square(h_0,h_\varepsilon) < 2 \delta_\square(\tilde{h}_0,\tilde{h}_\varepsilon)$, which can be done since $2 \delta_\square(\tilde{h}_0,\tilde{h}_\varepsilon)$ is strictly larger than $\inf_\sigma d_\square(h_0,h_\varepsilon^\sigma)$. Let $\tilde{q}_0 \in K^{-1}(\tilde{h}_0)$. Consider a coloring $c_0$ such that $q_0 = (h_0, c_0) \in \tau_\Theta^{-1}(\tilde{q}_0)$. Now define $q_\varepsilon \equiv (h_\varepsilon,c_0)$. Note that $\delta_\square^\Theta(\tilde{q}_0, \tilde{q}_\varepsilon) \le d_\square(h_0, h_\varepsilon) < 2 \delta_\square^\Theta(\tilde{h}_0, \tilde{h}_\varepsilon) < 2 \varepsilon$. Since $\tilde{q}_\varepsilon$ has the same coloring (in the sense of equivalence class) as $\tilde{q}_0$, we have that $\tilde{q}_\varepsilon \in \tilde{D}_{{\bf w}}$. Additionally, for $\varepsilon$ sufficiently small, since $\tilde{U}$ is open, we must have $\tilde{q}_\varepsilon \in \tilde{U}$, so $\tilde{q}_\varepsilon \in \tilde{U} \cap \tilde{D}_{\bf w}$. This implies $K(\tilde{q}_\varepsilon) = \tilde{h}_\varepsilon \in K(\tilde{U} \cap \tilde{D}_{\bf w})$, which is a contradiction. Therefore, we have that $K(\tilde{U} \cap \tilde{D}_{\bf w})$ is open, such that
    \begin{align}
        \inf_{\tilde{h} \in \textrm{int}(K(\tilde{U} \cap \tilde{D}_{{\bf w}}))} {\cal I}_p(\tilde{h}) = \inf_{\tilde{h} \in K(\tilde{U} \cap \tilde{D}_{{\bf w}})} {\cal I}_p(\tilde{h}) = \inf_{\tilde{q} \in \tilde{U}} {\cal J}^\Theta_{p,{\bf w}}(\tilde{q}).
    \end{align}
    This yields the lower bound for our large deviations principle:
    \begin{align}
        - \inf_{\tilde{q} \in \tilde{U}} {\cal J}^\Theta_{p,{\bf w}}(\tilde{q}) \le \liminf_{N \to \infty} \frac{1}{N^2} \log(\Tilde{\mathbb{P}}^\Theta_{N,p}(\tilde{U})).
    \end{align}

    Let $\tilde{F} \subseteq \tilde{{\cal W}}_\Theta$. To prove the upper bound, I will prove that $K(\tilde{F} \cap \tilde{D}_{{\bf w}})$ is closed. First, we can prove that $\tilde{D}_{{\bf w}}$ is closed. Let $\tilde{q}_0$ be a limit point of $\tilde{D}_{{\bf w}}$. That is, for all $\varepsilon > 0$, there is some $\tilde{q}_\varepsilon \in \tilde{D}_{{\bf w}}$ such that $\delta_\square^\Theta(\tilde{q}_0, \tilde{q}_\varepsilon) < \varepsilon$. Suppose $\tilde{q}_0 \notin \tilde{D}_{{\bf w}}$. Then we must have $\delta_{\Theta}(\tilde{c}_{q_0}, \tilde{c}_{{\bf w}}) > 0$. Now, we have that
    \begin{align}
        \varepsilon > \delta_\square^\Theta(\tilde{q}_0, \tilde{q}_\varepsilon) \ge \delta_{\Theta}(\tilde{c}_{q_0}, \tilde{c}_{q_\varepsilon}) = \delta_{\Theta}(\tilde{c}_{q_0}, \tilde{c}_{{\bf w}}).
    \end{align}
    For sufficiently small $\varepsilon$ this is a contradiction, so $\tilde{D}_{{\bf w}}$ is closed. Therefore, $\tilde{F} \cap \tilde{D}_{{\bf w}}$ is closed, and hence compact (since $\tilde{{\cal W}}_\Theta$ is compact). Now, we can prove that $K$ is a continuous function. Let $\varepsilon > 0$ and $\tilde{q} \in \tilde{{\cal W}}_\Theta$. Then for all $\tilde{q}'$ with $\delta_\square^\Theta(\tilde{q},\tilde{q}') < \varepsilon$ we have 
    \begin{align}
        \delta_\square(K(\tilde{q}), K(\tilde{q}')) = \delta_\square(\tilde{h}_q, \tilde{h}_{q'}) \delta_\square^\Theta(\tilde{q},\tilde{q}') < \varepsilon,
    \end{align}
    so $K$ is continuous. Since continuous functions map compact sets into compact sets, we have that $K(\tilde{F} \cap \tilde{D}_{{\bf w}})$ is compact, and hence closed by the generalized Heine-Borel theorem. Hence, we have that 
    \begin{align}
        \inf_{\tilde{h} \in \textrm{cl}(K(\tilde{F} \cap \tilde{D}_{{\bf w}}))} {\cal I}_p(\tilde{h}) = \inf_{\tilde{h} \in K(\tilde{F} \cap \tilde{D}_{{\bf w}})} {\cal I}_p(\tilde{h}) = \inf_{\tilde{q} \in \tilde{F}} {\cal J}^\Theta_{p,{\bf w}}(\tilde{q}).
    \end{align}
    This yields the desired upper bound:
    \begin{align}
        \limsup_{N \to \infty} \frac{1}{N^2} \log(\Tilde{\mathbb{P}}^\Theta_{N,p}(\tilde{F})) \le - \inf_{\tilde{q} \in \tilde{F}} {\cal J}^\Theta_{p,{\bf w}}(\tilde{q}),
    \end{align}
    which completes the proof.
\end{proof}

Using the previous result, we can state the convergence theorems that yield Theorem \ref{thm:mult_types_partition} in the paper.

\begin{theorem} \label{thm:colored_graphon_partition}
    Consider an Exponential Random Graph model where the set of node types is $\Theta$, where the measure is given by
    \begin{align}
        \pi_N(Q) = \exp\left\{ N^2 [{\cal U}(\Tilde{Q}) - \psi_N] \right\}.
    \end{align}
    If ${\cal U}$ is a continuous bounded function and the fraction of nodes of type $\theta$ converges to $w_\theta \in (0,1)$, then the normalization constant $\psi_N$ converges to
    \begin{align} \label{eq:colored_graphon_partition}
        \psi \equiv \sup_{\substack{\tilde{q} \in \Tilde{{\cal W}}_\Theta \\ \delta_\Theta(\tilde{c}_q,\tilde{c}_{{\bf w}}) = 0}} ({\cal U}(\Tilde{q}) + {\cal H}(\tilde{h}_q)).
    \end{align}
\end{theorem}

\begin{proof}
    This proof follows the same procedure as the proof of Theorem 3.1 in \citet{Chatterjee2013}, with the modifications made in \citet{Mele2017AFormation}. The $\delta_\Theta(\tilde{c}_q,\tilde{c}_{{\bf w}}) = 0$ constraint arises from the infinities in the rate function ${\cal J}_{{\bf w},1/2}^\Theta$.
\end{proof}

\begin{theorem} \label{thm:colored_graphon_convergence}
     Let $\Tilde{M}^*$ be the set of maximizers of the variational problem \eqref{eq:colored_graphon_partition}. Let $Q_N$ be a colored graph on $N$ vertices drawn from the model implied by function ${\cal U}$. Then, for any $\eta > 0$, there exist $C, \kappa > 0$ such that, for any $N$,
    \begin{align*}
        \mathbb{P}\{ \delta^\Theta_\square(\Tilde{Q}_N, \Tilde{M}^*) > \eta \} \le C e^{-N^2 \kappa},
    \end{align*}
    where $\mathbb{P}$ denotes the probability measure implied by the model.
\end{theorem}

\begin{proof}
    This proof follows the same procedure as the proof of Theorem 3.2 in \citet{Mele2017AFormation}.
\end{proof}

\end{document}